\documentclass[reprint,twocolumn,superscriptaddress,showpacs]{revtex4-1}

\usepackage{amsmath,amssymb,amsfonts}
\usepackage{bm}
\usepackage{braket}

\usepackage[dvipdfmx]{graphicx}
\usepackage{amsmath,bbm,amssymb,bm,times}
\usepackage[normalem]{ulem}
\usepackage[usenames,dvipsnames]{color}
\definecolor{dgreen}{rgb}{0.0, 0.5, 0.0}

\begin{document}

\title{Rectification in Nonequilibrium Steady States of Open Many-Body Systems}
\author{Kazuki Yamamoto}
\email{yamamoto.kazuki.72n@st.kyoto-u.ac.jp}
\affiliation{Department of Physics, Kyoto University, Kyoto 606-8502, Japan}
\author{Yuto Ashida}
\affiliation{Department of Applied Physics, University of Tokyo, Tokyo 113-8656, Japan}
\affiliation{Department of Physics, University of Tokyo, Hongo, Tokyo 113-0033, Japan}
\affiliation{Institute for Physics of Intelligence, University of Tokyo, Hongo, Tokyo 113-0033, Japan}
\author{Norio Kawakami}
\affiliation{Department of Physics, Kyoto University, Kyoto 606-8502, Japan}

\date{\today}

\begin{abstract}
We study how translationally invariant couplings of many-particle systems and nonequilibrium baths can be used to rectify particle currents, for which we consider minimal setups to realize bath-induced currents in nonequilibrium steady states of one-dimensional open fermionic systems.  We first analyze dissipative dynamics associated with a nonreciprocal Lindblad operator and identify a class of Lindblad operators that are sufficient to acquire a unidirectional current. We show that unidirectional particle transport can in general occur when a Lindblad operator is reciprocal provided that the inversion symmetry and the time-reversal symmetry of the microscopic Hamiltonian are broken. We demonstrate this mechanism on the basis of both analytical and numerical approaches including the Rashba spin-orbit coupling and the Zeeman magnetic field.
\end{abstract}

\maketitle

%%%-----[Introduction]-----
\section{Introduction}
In recent years, open quantum systems are widely explored as exemplified by driven-dissipative many-body systems \cite{Diehl08, Kraus08, Muller12, Daley14, Ikeda20} and non-Hermitian phenomena \cite{Ashida20}. They have revealed that dissipation can qualitatively change various aspects of many-body physics such as in quantum critical phenomena \cite{Sieberer13, Ashida17, Ashida16, Nakagawa18}, phase transitions \cite{Diehl10, Honing12, Hamazaki19, Yamamoto19, Matsumoto19}, magnetism \cite{Diehl10a, Yi12, Nakagawa20}, and quench dynamics \cite{Durr09, Ashida18, Yamamoto20}. In particular, experimental advances in controlling dissipation have allowed one to study nonequilibrium and non-Hermitian phenomena in trapped ions \cite{Barreiro11, Schindler13}, photonics \cite{Liu11, Piilo18}, ultracold atoms \cite{Ott13, Ott15, Tomita17, Spon18, Tomita19, Gerbier19, Takasu20}, and exciton-polariton systems \cite{Kasprzak06, Leyder07, Byrnes14, Gao15, Baboux16, Klembt18}. These remarkable developments have offered new opportunities for exploring intriguing phenomena unique to open quantum systems in {\it homogeneous} setups in contrast to, e.g., boundary-driven systems \cite{William90}.

On another front, nonreciprocal phenomena, which have been a long-standing problem in condensed matter physics and nonequilibrium statistical mechanics, play a vital role in a variety of areas, including solid-state physics \cite{Tokura18, Rikken97, Linke99, Morimoto16, Kitamura20, Ono02, Molenkamp08}, photonics \cite{Hamidreza10, Chang14, Peng14, Nazari14, Sylvain15}, acoustics \cite{Liang09, Andrea14, Li14, Chen15, Daraio18, Ruzzene19}, and active matter \cite{Estep14, Coulais17, Coulais19, Liao20}. While p-n junctions are nonreciprocal devices of commercial success, there is significant interest in exploring alternative mechanisms, and recent discoveries have shed light on generating nonreciprocal flows without any temperature biases \cite{Flach02, Souvik02, Ren10, Zhu16, Sabass17, Liao20}. While dissipation has been recognized as a key ingredient to control transport properties, Onsager's reciprocal theorem \cite{Onsager31, Benenti17} prohibits rectification by equilibrium baths and thus it is of central importance to introduce  {\it nonequilibrium} baths.

In open quantum systems, one common way to introduce rectification is to couple a system with two different baths at  boundaries and use temperature gradients as exemplified by thermal diodes \cite{Werlang14, Pereira17b, Thomas18, Pereira19, Pereira19E, Andreu19, Balachandran19}. Indeed, many of the previous studies have focused on inhomogeneous setups such as by introducing boundary driving \cite{Saito03, Hannu07, Prosen11, Landi14, Zala15, Pereira17a, Pereira18, Lucas16, Balachandran18, Daniel18, Karen19, Eduardo19, Damanet19}. In contrast, rectification induced by \textit{homogeneous} dissipation of nonequilibrium baths has scarcely been explored despite recent experimental advances mentioned above. To our knowledge, there are so far only a few studies in this direction, where nonreciprocal photon transmissions \cite{Metelmann15, Lodahl17, Keck18} and rectified heat currents in spin chains \cite{Zala18a} are discussed. Thus, it is still unclear how translationally invariant homogeneous dissipation of nonequilibrium baths can be harnessed to realize unidirectional fermionic transports.
\begin{figure}[b]
\includegraphics[width=8.5cm]{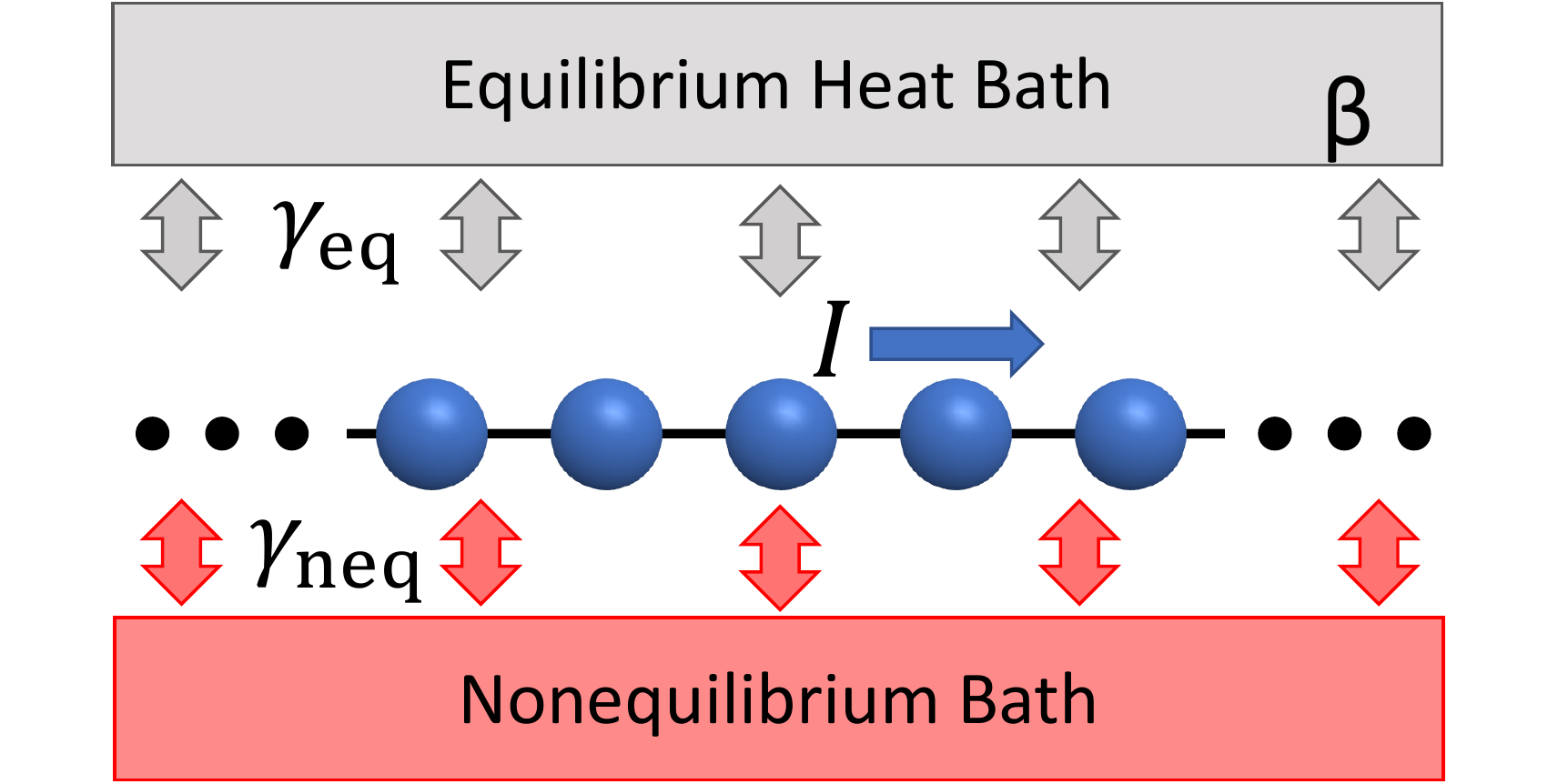}
\caption{Schematic illustration of our setup. Fermions are trapped in a one-dimensional lattice and uniformly coupled to a nonequilibrium bath, which gives rise to translationally invariant dissipation. An equilibrium heat bath with the inverse temperature $\beta$ is also coupled to the system to ensure that the system reaches to the Gibbs state in the absence of nonequilibrium driving. The coupling strength to each bath is given by $\gamma_{\mathrm {eq}}$ and $\gamma_{\mathrm{neq}}=1-\gamma_{\mathrm{eq}}$, respectively. Unidirectional current $I$ can arise in NESS only when the system is driven out of equilibrium.}
\label{fig1}
\end{figure}

In this paper, we propose minimal setups to obtain a unidirectional particle transport in nonequilibrium steady states (NESS) of one-dimensional open fermionic systems, where a nonequilibrium bath is uniformly coupled to the system and gives rise to homogeneous dissipation (see Fig.~\ref{fig1}). We first consider a nonreciprocal Lindblad operator, which is translationally invariant and conserves the particle number of the system, and elucidate a general condition to acquire a nonreciprocal particle transport in NESS. We numerically calculate the current by considering a specific dissipator that can be realized in ultracold atoms \cite{Diehl08, Kraus08}. Then, we demonstrate that a reciprocal Lindblad operator can also induce unidirectional particle transport in NESS provided that the inversion symmetry and the time-reversal symmetry of the Hamiltonian are broken. We consider spin-dependent dephasing as a reciprocal Lindblad operator and evaluate the current by analytical and numerical methods in the presence of the Rashba spin-orbit coupling and the Zeeman magnetic field \cite{Birkholz08}. Our results should be tested by using ultracold atoms or semiconductor quantum dots, where the master-equation description can be used \cite{Nakajima18}.

%%%-----[model]------
\section{Model}
We consider a one-dimensional lattice model coupled to both an equilibrium heat bath and a nonequilibrium Markovian bath. Such a situation is described by the Lindblad master equation
\begin{gather}
\partial_t \rho = -i[H_0, \rho] +\mathcal L_1 \rho,\\
\mathcal L_1 \rho= \epsilon(\gamma_{\mathrm{eq}}\mathcal D^{\mathrm{eq}}(\rho) + \gamma_{\mathrm{neq}}\mathcal D^{\mathrm{neq}}(\rho)),
\end{gather}
with dissipators
\begin{align}
\mathcal D^{(i)}(\rho) =\sum_m\left(L_m^{(i)}\rho L_m^{(i)\dag}-\frac{1}{2}\left\{L_m^{(i)\dag}L_m^{(i)}, \rho\right\}\right),
\end{align}
where $H_0$ is a noninteracting Hamiltonian governing the internal dynamics, $L_m$ is a so-called Lindblad operator, $\gamma_{\mathrm{eq}}$ and $\gamma_{\mathrm{neq}}=1-\gamma_{\mathrm{eq}}$ denote the relative coupling strengths between two baths, $\gamma_{\mathrm{eq}}\in[0, 1]$. We  assume that the baths are weakly coupled to the system with a small dimensionless parameter $\epsilon$. Here and henceforth, we set $\hbar=1$. In AMO systems, the approximations involved in deriving the Lindblad master equation are typically well-satisfied to many orders of magnitudes, and a lot of experimental studies have revealed that these approximations are indeed applicable to various situations \cite{Daley14}. The Lindblad equation can be derived from a fully microscopic Hamiltonian of the system, the system-bath coupling, and the bath after tracing out the bath degrees of freedom with the Born approximation, Markov approximation, and rotating-wave approximation. We remark that the present model is an intrinsically interacting many-body problem because the dissipator cannot in general be expressed in terms of quadratic annihilation/creation operators as detailed below.

\section{Time-dependent generalized GIbbs ensembles}
When the integrability of the translationally invariant internal system is weakly broken due to the coupling with the reservoir, the time evolution of the system can be described by a time-dependent generalized Gibbs ensemble (tGGE) \cite{Vidmar16, Zala17, Zala18a, Zala18b}, which is justified for times $t$ of the order of $1/\epsilon$ and larger,
\begin{align}
\rho_{\mathrm {GGE}}(t) = \frac{e^{-\sum_q \lambda_q(t)I_q}}{\mathrm{Tr}[e^{-\sum_q \lambda_q(t)I_q}]},
\label{eq_GGE}
\end{align}
where $I_q$ is an approximately conserved quantity as a consequence of weak driving. Previous studies \cite{Zala17, Zala18a, Zala18b} have shown that, by applying a linear-order perturbation theory to the Lindblad equation, one can obtain a differential equation that determines the dynamics of Lagrange parameters
\begin{gather}
\dot \lambda_q =-\sum_p(\chi(t)^{-1})_{qp}\mathrm{tr}\left[I_p\mathcal L_1\rho_{\mathrm{GGE}}(t)\right],\label{eq_rate}\\
\chi_{qp}(t)=\langle I_qI_p\rangle_{\mathrm{GGE}}-\langle I_q\rangle_{\mathrm{GGE}}\langle I_p\rangle_{\mathrm{GGE}},
\label{eq_chi}
\end{gather}
where $\langle\cdots\rangle_{\mathrm{GGE}}=\mathrm{tr}[\cdots\rho_{\mathrm{GGE}}(t)]$ (for the detailed calculation, see Appendix~\ref{sec_gge}). We note that $\lambda_q$ and $\langle I_q \rangle_{\mathrm{GGE}}$ are of the order of $\epsilon^0$ in NESS, and this fact causes large current responses at arbitrarily weak system-bath coupling as shown below. The validity of these equations has been shown in Ref.~\cite{Zala18a} by comparing the results obtained from tGGE with those from the exact diagonalization, as also seen from Fig.~\ref{fig5} below.

%%%-----[result]------
\section{Results}
In the following, we evaluate the current in NESS by using tGGE approach with Eqs.~\eqref{eq_GGE}--\eqref{eq_chi} at arbitrarily weak system-bath coupling. We propose two minimal setups for rectifying the current in NESS both for nonreciprocal dissipator and reciprocal dissipator.
\subsection{Rectification by nonreciprocal dissipator}
We first consider the one-dimensional tight-binding model
\begin{align}
H_0 = -J\sum_{j=0}^{L-1}(c_{j+1}^\dag c_j + \mathrm{H.c.})=\sum_{-\pi\le k<\pi}\epsilon_k c_k^\dag c_k,
\end{align}
where $J$ is the hopping amplitude and $\epsilon_k=-2J\cos (k)$ is the eigenspectrum. We focus on the homogeneous couplings with nonequilibrium baths of infinite system sizes and assume that the system is subject to periodic boundary conditions and periodic dissipation of length $L$. Here, we note that a realistic system is sometimes affected by a particle source and sink at the edges, but we ignore such effects for simplicity. Then, $I_q$ in Eq.~\eqref{eq_GGE} is given by the local number operator in the momentum space $I_q=c_q^\dag c_q$.

\begin{figure}[b]
\includegraphics[width=8.5cm]{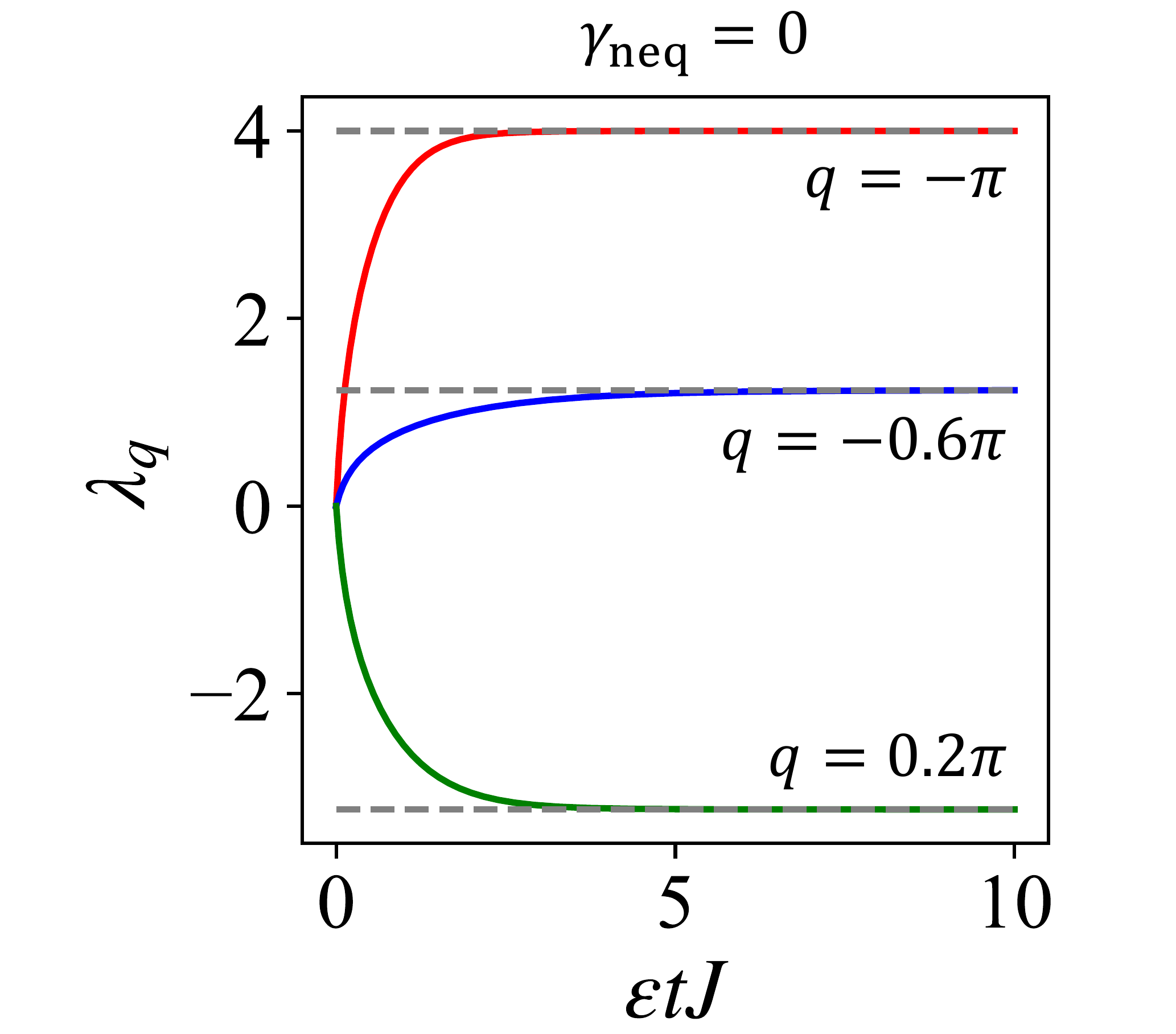}
\caption{Dynamics of Langrange parameters $\lambda_q$ without nonequilibrium driving obtained from Eq.~\eqref{eq_rate1}, which satisfies the detailed balance condition Eq.~\eqref{eq_detailed1} with $\gamma_{\mathrm{eq}}=1$. The system goes to the Gibbs state (dashed lines) after sufficiently long-time evolution of the order of $1/\epsilon$. The initial state is set to infinite temperature. The parameters are set to $\beta=2/J$ and $\epsilon=0.05$.}
\label{fig5}
\end{figure}

\begin{figure}[b]
\includegraphics[width=8.5cm]{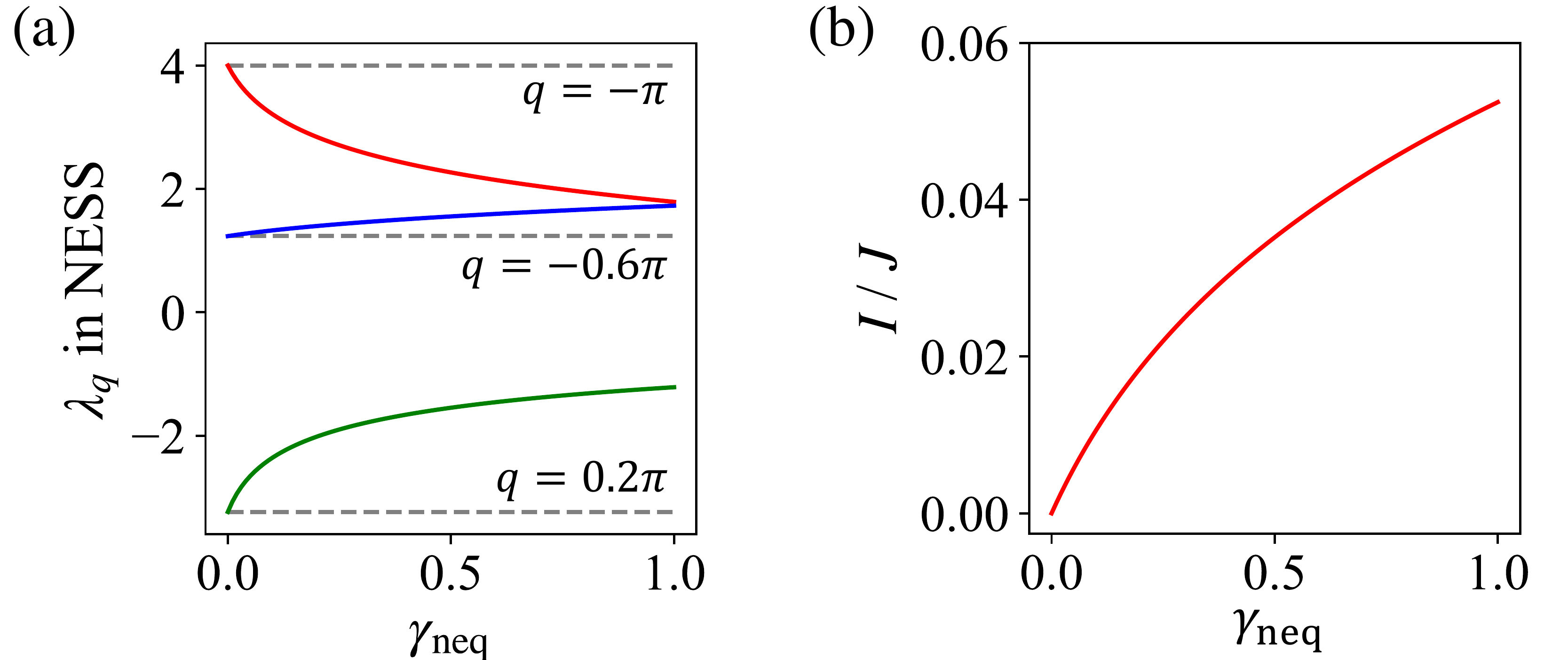}
\caption{(a) Lagrange parameters in NESS that are driven out of equilibrium as the nonequilibrium dissipation rate $\gamma_{\mathrm{neq}}$ is increased. The grey dashed line denotes the Gibbs state. (b) Current $I$ in NESS as a function of $\gamma_{\mathrm{neq}}$ with Lindblad operators~\eqref{eq_detailed1} and \eqref{eq_nonrecipro}. The parameters are set to $\beta=2/J$, $\epsilon=0.05$, $\delta=1+i$, and $\delta^\prime=1+0.5i$.}
\label{fig2}
\end{figure}

The Lindblad operators corresponding to the equilibrium heat bath satisfy $[L_m, H_0]=\zeta_m L_m$ with $\zeta_m= \epsilon_k - \epsilon_l$, $m=(k, l)\in \{-\pi, -\pi+2\pi/N, ...,\pi-2\pi/N\}$ to ensure the detailed balance condition $L_{kl}^\dag=L_{lk}e^{-\beta(\epsilon_k-\epsilon_l)/2}$ \cite{Liu14, Gong16, Ikeda20} in such a way that, without nonequilibrium driving, the system goes to the Gibbs state $\rho_{\mathrm{can}}=e^{-\beta H_0}/\mathrm{tr}(e^{-\beta H_0})$ irrespective of the initial state [see Figs.~\ref{fig5} and \ref{fig2}(a)]. For the sake of simplicity, we here employ the following Lindblad operator corresponding to the equilibrium heat bath
\begin{align}
L_{lk}^{\mathrm{eq}} = \sqrt \frac{J}{L}c_l^\dag c_k e^{\beta(\epsilon_k - \epsilon_l)/4}.
\label{eq_detailed1}
\end{align}

To realize current rectification in NESS, we consider a nonreciprocal Lindblad operator corresponding to the nonequilibrium bath and assume that it is translationally invariant and conserves the particle number of the system. In this case, the Lindblad operator can in general be labeled by a wave number with coefficients $\Delta_{kq}$ as
\begin{align}
L^{\mathrm{neq}}_q=\sqrt \frac{J}{L}\sum_{-\pi\le k<\pi} \Delta_{kq} c_{k-q}^\dag c_k.
\label{eq_dissipation1}
\end{align}
 Using Eq.~\eqref{eq_rate} and Lindblad operators \eqref{eq_detailed1} and \eqref{eq_dissipation1}, we obtain the rate equation that governs the dynamics of the system (see Appendix \ref{sec_rate} for detailed derivations)
\begin{align}
\dot \lambda_q=-\frac{\epsilon J}{L}\frac{1+e^{-\lambda_q}}{e^{-\lambda_q}}(\gamma_{\mathrm{eq}}F_q^{\mathrm{eq}}+\gamma_{\mathrm{neq}}F_q^{\mathrm{neq}}),
\label{eq_rate1}
\end{align}
where
\begin{gather}
F_q^{\mathrm{eq}}=\sum_{-\pi\le k<\pi}\frac{e^{\beta(\epsilon_k-\epsilon_q)/2-\lambda_k}-e^{\beta(\epsilon_q-\epsilon_k)/2-\lambda_q}}{1+e^{-\lambda_k}},\displaybreak[2]\\
F_q^{\mathrm{neq}}=\sum_{-\pi\le k<\pi}\frac{|\Delta_{k, k-q}|^2e^{-\lambda_k}-|\Delta_{q,q-k}|^2e^{-\lambda_q}}{1+e^{-\lambda_{k}}}.\label{eq_Fneq1}
\end{gather}

We numerically solve the rate equation \eqref{eq_rate1} to obtain the dynamics of Lagrange parameters and their steady-state values. We first verify that Lagrange parameters go to the Gibbs state in NESS if there is no nonequilibrium driving. Figure~\ref{fig5} shows the relaxation dynamics of Lagrange parameters, which obey Eq.~\eqref{eq_rate1} with $\gamma_{\mathrm{eq}}=1$ satisfying the detailed balance condition Eq.~\eqref{eq_detailed1}. We see that the system goes to the Gibbs state (grey dashed lines) after sufficiently long time evolution. Then, we calculate steady-state values of Lagrange parameters following the rate equation \eqref{eq_rate1}. We see that Lagrange parameters depart from the Gibbs state when the system is driven out of equilibrium as the nonequilibrium dissipation rate $\gamma_{\mathrm{neq}}$ is increased [see Fig.~\ref{fig2}(a)].

We now derive a general condition to realize a nonzero nonreciprocal current in NESS. The current $I$ generally consists of two terms including Hamiltonian current and dissipative current of order $\epsilon$. For such a small $\epsilon$ that justifies the tGGE approach, the dissipative current can be ignored, which is consistent with a general description of the current in open quantum systems \cite{Gebauer04, Bodor06}. We obtain the current from the continuity equation for the density matrix as $I=2/L\sum_j\mathrm{Im}\langle c_j^\dag c_{j-1}(H_0)_{j,j-1}\rangle_{\mathrm{GGE}}$ \cite{Erasmo92, Mahan00}, where $(H_0)_{j,j-1}$ denotes the coefficient of $c_j^\dag c_{j-1}$ in $H_0$. In the present model, the current, which is the order of $\epsilon^0$, is given by
\begin{align}
I&=\frac{iJ}{L}\sum_{j=0}^{L-1} \langle c_{j+1}^\dag c_j - \mathrm{H.c.}\rangle_{\mathrm{GGE}}\notag\\
&=\frac{2J}{L}\sum_{-\pi\le q<\pi} \sin (q)\frac{e^{-\lambda_q}}{1+e^{-\lambda_q}}.
\end{align}
Thus, to obtain a nonreciprocal current, the Lagrange parameter $\lambda_q$ must not be an even function of $q$. More specifically,  as inferred from Eq.~\eqref{eq_Fneq1}, this condition requires a set of $(k$, $q)\in[-\pi,$ $\pi)$ to satisfy (at least) one of the following conditions:
\begin{align}
|\Delta_{q,k+q}|\neq|\Delta_{-q, k-q}|,
\quad|\Delta_{k, k-q}|\neq|\Delta_{k, k+q}|.
\label{eq_condition}
\end{align}
We note that the time-reversal symmetry of the internal Hamiltonian is not broken, and an even function $\lambda_q$ prohibits the rectification of the current even when the dissipative current of order $\epsilon$ is included because it leads to a parity-even distribution of particles in real space and thus the current (that is parity-odd) cannot exist.

Let us apply the condition~\eqref{eq_condition} for obtaining the nonreciprocal current to a specific example. We introduce a phenomenological dissipator which is proposed in ultracold atoms in an optical lattice illuminated by Raman laser \cite{Diehl08, Kraus08},
\begin{align}
L_j^{\mathrm{neq}}=\sqrt J(c_j^\dag+\delta c_{j+1}^\dag)(c_j-\delta^\prime c_{j+1}),
\label{eq_nonrecipro}
\end{align}
where the subscript $j$ denotes the lattice site. This type of Lindblad operator causes the enhancement or suppression of the atomic phases in two adjacent lattice sites and it is not obvious how such type of superposition of the atomic phases leads to the transport of atoms. We rewrite Eq.~\eqref{eq_nonrecipro} as
\begin{align}
L_q^{\mathrm{neq}}=\sqrt \frac{J}{L}\sum_k(1+\delta e^{-i(k-q)})(1-\delta^\prime e^{ik})c_{k-q}^\dag c_k,
\label{eq_nonreciprok}
\end{align}
where we set the lattice constant $a=1$. From Eq.~\eqref{eq_condition}, the Lindblad operator \eqref{eq_nonreciprok} should give rise to a nonreciprocal current when either $\delta$ or $\delta^\prime$ has the imaginary part. This is demonstrated in Fig.~\ref{fig2}(b), where the current in NESS is plotted as a function of $\gamma_{\mathrm{neq}}$ for $\delta=1+i$, $\delta^\prime=1+0.5i$. We see that a large current is built up on a timescale of $1/\epsilon$ as it is driven out of equilibrium though it exactly vanishes in equilibrium ($\gamma_{\mathrm{neq}}=0$).

\subsection{Rectification by reciprocal dissipator}
\begin{figure}[b]
\includegraphics[width=8.5cm]{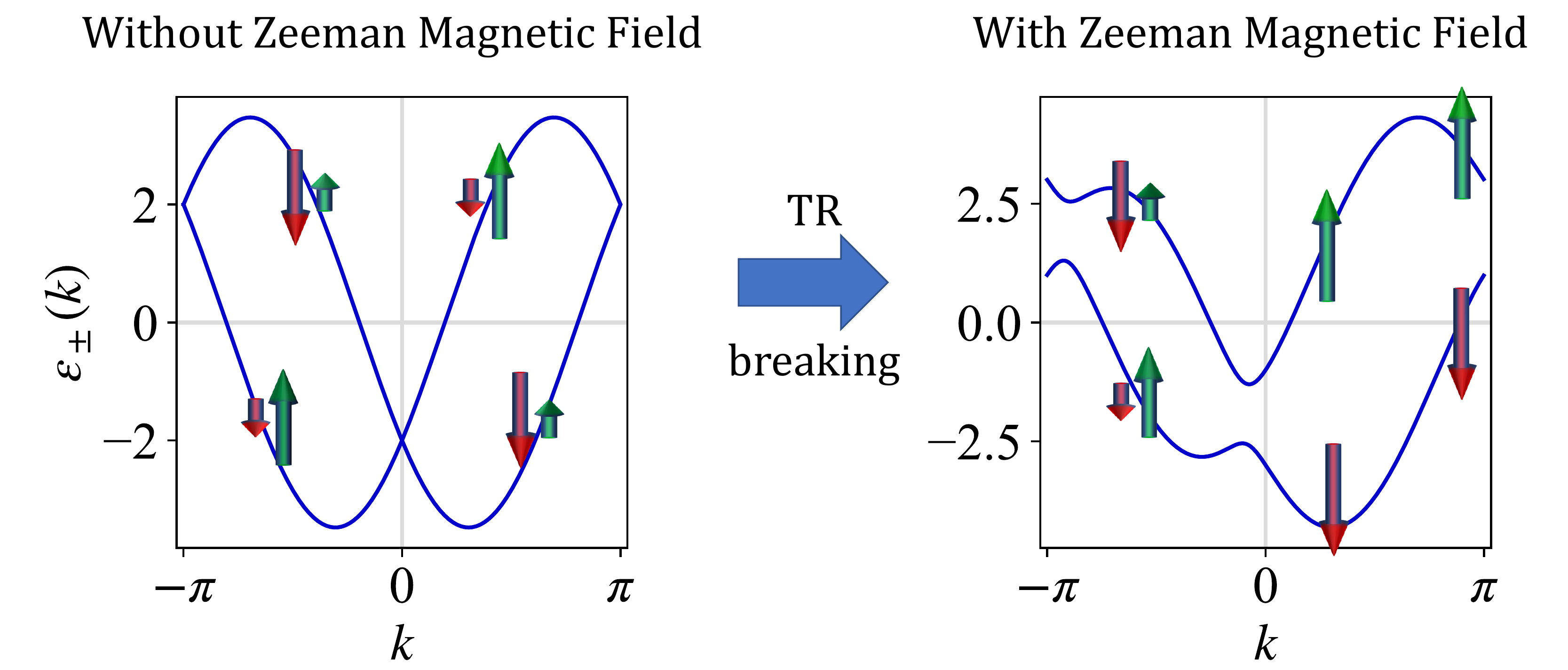}
\caption{Schematic illustration of the energy spectrum of the tight-binding Hamiltonian with Rashba spin-orbit coupling Eq.~\eqref{eq_Rashba}. The time-reversal symmetry (TR) of the Hamiltonian is broken when the Zeeman magnetic field is applied to the system.}
\label{fig3}
\end{figure}

We next discuss how to realize a nonzero nonreciprocal current by a {\it reciprocal} Lindblad operator at the expense of the broken inversion and time-reversal symmetries of the internal Hamiltonian. To be concrete, we include the Rashba spin-orbit coupling and the Zeeman magnetic field into the one-dimensional tight-binding model \cite{Birkholz08}
\begin{align}
H_0 = &-J\sum_{j\sigma}(c_{j+1\sigma}^\dag c_{j\sigma} + \mathrm{H.c.}) + h\sum_{j=0}^{L-1}(n_{j\uparrow}-n_{j\downarrow}) \notag\\
&-\alpha_z\sum_{j\sigma\sigma^\prime}(c_{j+1\sigma}^\dag(i\sigma_y)_{\sigma\sigma^\prime} c_{j\sigma^\prime} + \mathrm{H.c.})\notag\\
&+ \alpha_y\sum_{j\sigma\sigma^\prime}(c_{j+1\sigma}^\dag(i\sigma_z)_{\sigma\sigma^\prime} c_{j\sigma^\prime} + \mathrm{H.c.})\notag\\
=&\sum_{-\pi\le k<\pi}\sum_{\nu=\pm}\epsilon_{k\nu}\eta_{k\nu}^\dag\eta_{k\nu},
\label{eq_Rashba}
\end{align}
where $h$ denotes the Zeeman splitting, $\sigma_{y,z}$ are the Pauli matrices, $\alpha_{y,z}$ denote the Rashba hopping with spin flips, $\sigma=\uparrow\downarrow$ and $\nu=\pm$ label spin and band indices, respectively, and the system is subject to periodic boundary conditions and periodic dissipation of length $L$. The Rashba spin-orbit coupling and the Zeeman magnetic field break the inversion symmetry and the time-reversal symmetry of the Hamiltonian, respectively (see Fig.~\ref{fig3}). The Hamiltonian is diagonalized with eigenvalues $\epsilon_{k\pm}=-2J\cos (k)\pm\sqrt{(2\alpha_y \sin (k)+h)^2 +4\alpha_z^2\sin^2 (k)}$ and quasiparticle operators $\eta_{k\pm}$, which are given by a unitary transformation as $c_{k\sigma}=\sum_\nu u_{\sigma \nu}(k)\eta_{k \nu}$ and obey the anticommutation relation $\{\eta_{k\mu},$ $\eta_{k^\prime\nu}^\dag\}=\delta_{kk^\prime}\delta_{\mu\nu}$ (see Appendix \ref{sec_quasi} for details). In this case, local conservation laws of few-body observables are given by the number operators of quasiparticles $I_{q\nu}=\eta_{q\nu}^\dag\eta_{q\nu}$ [cf. Eq.~\eqref{eq_GGE}].

To identify the Lindblad operators $L_m^{\mathrm{eq}}$ that satisfy the detailed balance condition, we consider $\mu(\nu)$ dependence for the energy bands of quasiparticles in addition to Eq.~\eqref{eq_detailed1}:
\begin{align}
L_{l\mu, k\nu}^{\mathrm{eq}} = \sqrt \frac{J}{L} \eta_{l\mu}^\dag \eta_{k\nu} e^{\beta(\epsilon_{k\nu} - \epsilon_{l\mu})/4}.
\label{eq_detailed2}
\end{align}
The relaxation dynamics of Lagrange parameters, which follow the detailed balance condition \eqref{eq_detailed2}, is qualitatively the same as that in Fig.~\ref{fig5} except the fact that the degrees of freedom are doubled.
As the reciprocal Lindblad operator of the nonequilibrium bath, we consider the spin-dependent dephasing given by
\begin{align}
L_{j\sigma}^{\mathrm{neq}}=\sqrt{J\gamma_\sigma}c_{j\sigma}^\dag c_{j\sigma},
\label{eq_dissipation2}
\end{align}
where $j$ labels the lattice site and the dissipation rates of up and down spins satisfy $\gamma_\uparrow +\gamma_\downarrow=1$. We calculate the rate equation for the Lagrange parameters~\eqref{eq_rate} with Lindblad operators \eqref{eq_detailed2} and \eqref{eq_dissipation2}, which is given by (see Appendix \ref{sec_rate} for detailed calculations)
\begin{align}
\dot \lambda_{q\nu}=-\frac{\epsilon J}{L}\frac{1+e^{-\lambda_{q\nu}}}{e^{-\lambda_{q\nu}}}(\gamma_{\mathrm{eq}}F_{q\nu}^{\mathrm{eq}}+\gamma_{\mathrm{neq}}F_{q\nu}^{\mathrm{neq}})
\label{eq_rate2}
\end{align}
with the force
\begin{gather}
F_{q\nu}^{\mathrm{eq}}=\sum_{k\mu}\frac{e^{\beta(\epsilon_{k\mu}-\epsilon_{q\nu})/2-\lambda_{k\mu}}-e^{\beta(\epsilon_{q\nu}-\epsilon_{k\mu})/2-\lambda_{q\nu}}}{1+e^{-\lambda_{k\mu}}},\\
F_{q\nu}^{\mathrm{neq}}=\sum_{k \mu \sigma}\gamma_\sigma|u_{\sigma\nu}(q)|^2|u_{\sigma\mu}(k)|^2\frac{e^{-\lambda_{k\mu}}-e^{-\lambda_{q\nu}}}{1+e^{-\lambda_{k\mu}}}.\label{eq_Fneq2}
\end{gather}
We see from Eq.~\eqref{eq_Fneq2} that the system goes to the infinite temperature state, i.e.,  $\lambda_q=0$ for all $q$, without equilibrium heat bath. Nevertheless, the current can rectify if the system couples to both equilibrium and nonequilibrium baths.

When the dynamics is determined from the rate equation \eqref{eq_rate2}, the current which is the order of $\epsilon^0$ can be obtained from the continuity equation for the density matrix as
\begin{align}
I = \sum_{\sigma=\uparrow\downarrow}I_\sigma,
\label{eq_currenttotal}
\end{align}
where the spin-resolved current $I_\sigma$, is given by (see Appendix \ref{sec_quasi} for details)
\begin{align}
I_\sigma =&-\frac{i}{L}\bigg[-J\sum_j\langle c_{j+1\sigma}^\dag c_{j\sigma} - \mathrm{H.c.}\rangle_{\mathrm{GGE}}\notag\\
&-\alpha_z\sum_{j\sigma^\prime}\langle c_{j+1\sigma}^\dag(i\sigma_y)_{\sigma\sigma^\prime} c_{j\sigma^\prime} - \mathrm{H.c.}\rangle_{\mathrm{GGE}}\notag\\
&+ \alpha_y\sum_{j\sigma^\prime}\langle c_{j+1\sigma}^\dag(i\sigma_z)_{\sigma\sigma^\prime} c_{j\sigma^\prime} - \mathrm{H.c.}\rangle_{\mathrm{GGE}}\bigg].
\label{eq_current2}
\end{align}
\begin{figure}[b]
\includegraphics[width=8.5cm]{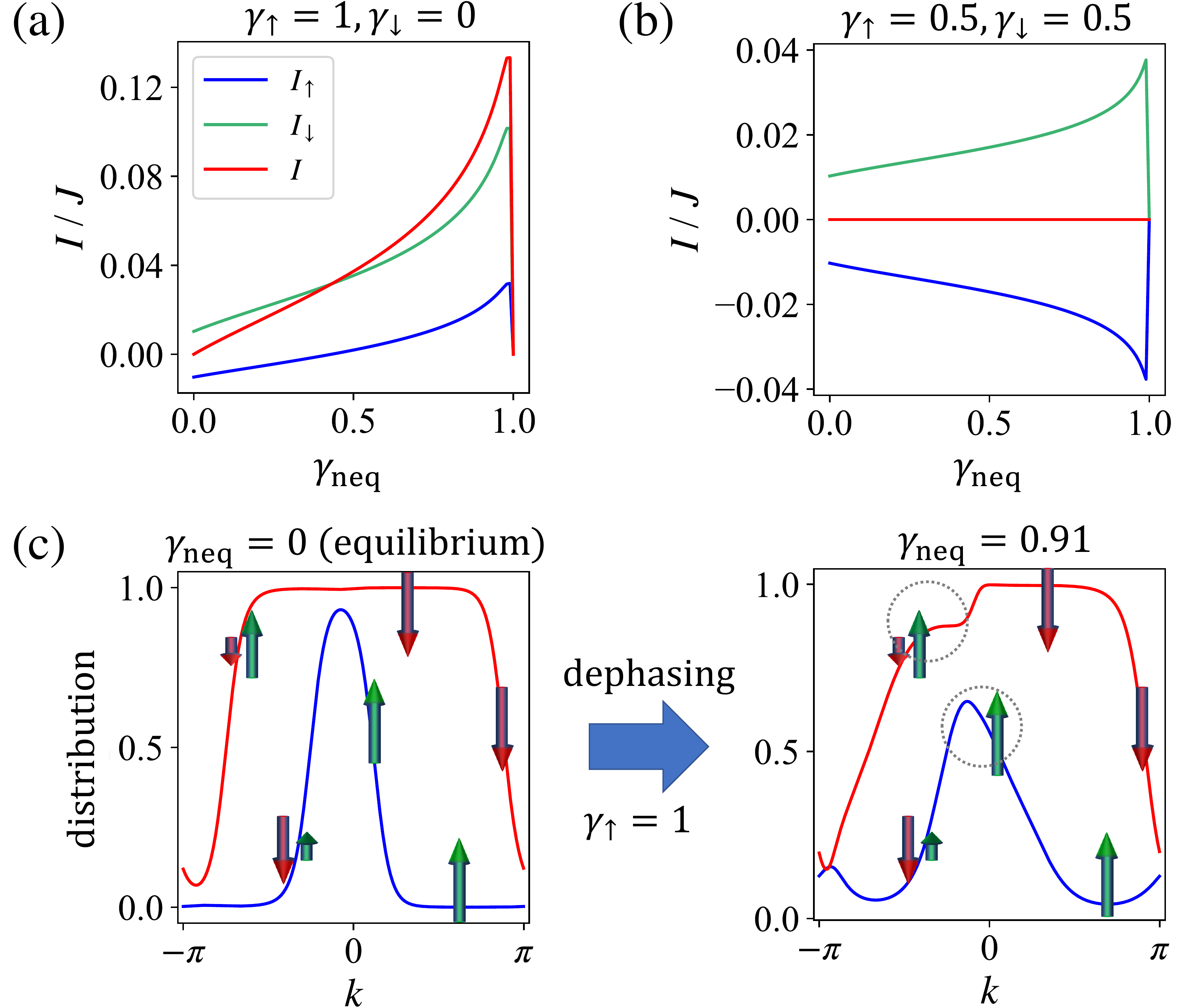}
\caption{(a,b) NESS current and its spin dependence as a function of $\gamma_{\mathrm{neq}}$ in the presence of the Zeeman magnetic field and the Rashba spin-orbit coupling. Dephasing is applied to up spins in (a), and to both up and down spins with equal rates in (b). (c) Distribution of the upper band (blue) and the lower band (red) in NESS for the equilibrium Gibbs state (left) and the nonequilibrium state where dephasing is applied to up spins (right). Population changes are enhanced near the Fermi surface due to dephasing (marked by grey dotted circles). The parameters are set to $\beta=2/J$, $\epsilon=0.05$, $\alpha_y=1.1J$, $\alpha_z=0.9J$, and $h=J$. The initial state is at infinite temperature.}
\label{fig4}
\end{figure}
We have confirmed that, by numerical calculations using Eq.~\eqref{eq_currenttotal}, the current $I$ is nonzero only when both the Zeeman magnetic field and the Rashba spin-orbit coupling exist. This can be understood as follows. Since dissipation by an equilibrium bath does not rectify the current, one must resort to a nonequilibrium bath for obtaining a nonzero nonreciprocal current. From Eq.~\eqref{eq_Fneq2}, we see that nonreciprocity of the distribution of Lagrange parameters is determined from the property of the unitary transformation of quasiparticles, namely, the symmetry of the internal Hamiltonian $H_0$. In fact, due to the structure of the matrix component $u_{\sigma\nu}(k)$ (see Appendix~\ref{sec_quasi}), dephasing by the nonequilibrium bath in Eq.~\eqref{eq_Fneq2} contributes to the Lagrange parameters as an even function with respect to $q$ if either one of the Zeeman magnetic field or the Rashba spin-orbit coupling is absent. As inversion-symmetric Lagrange parameters give the parity-even distribution of particles in real space, the current does not rectify even if dissipative correction of the order of $\epsilon$ is included.

Figures~\ref{fig4}(a) and (b) show the currents in NESS in the presence of the Rashba spin-orbit coupling and the Zeeman magnetic field. As shown in Fig.~\ref{fig4}(a), the dephasing applied to up spins leads to a large nonreciprocal current $I$ in NESS after a time evolution set by $1/\epsilon$, and it becomes larger as the system is driven out of equilibrium. We recall that the system goes to the Gibbs state for $\gamma_{\mathrm{neq}}=0$ and the infinite temperature state for $\gamma_{\mathrm{neq}}=1$, both of which do not rectify the total current $I$. When the dephasing is applied to both up and down spins with equal rates [see Fig.~\ref{fig4}(b)], the total current $I$ vanishes irrespective of the dissipation rate $\gamma_{\mathrm{neq}}$, as up spins and down spins contribute to the current in the opposite directions and cancel out (see Appendix~\ref{sec_downspin}). Here, we note that the sharp peak of the current in Fig.~\ref{fig4}(a) comes from the sudden heating up to the infinite temperature due to the nonequilibrium bath and the peak position can be controlled by the system parameters, e.g., the Zeeman magnetic field $h$.

Physically, rectification of the current in NESS can be understood from the change of spin distribution near the Fermi surface. As shown in the left panel of Fig.~\ref{fig4}(c), the spin distribution forms an effective Fermi surface in the steady state [see also the right panel in Fig.~\ref{fig3}], reflecting the half-filled initial state. When dephasing is applied to up spins [see the right panel in Fig.~\ref{fig4}(c)], they heat up and spins near the Fermi surface are most likely to move to the other eigenstates. As a result, the number of particles near the Fermi surface where up spins exist decreases, thereby contributing to the current in the positive direction [see also Eqs.~\eqref{eq_currentS1} and \eqref{eq_currentS2} in Appendix~\ref{sec_quasi}]. However, as shown in Fig. 4, the main contribution to the current originates from down spins where dephasing is not applied, because up spins heated up by dephasing move to cancel out the contribution to the current.

%%%-----[Conclusion]-----
\section{Discussions}We demonstrate arbitrarily weak translationally invariant system-bath coupling can induce large rectification in \textit{homogeneous} open quantum systems, which arises from the interplay between nonequilibrium dissipator and internal Hamiltonian dynamics. This contrasts with conventional setups in, e.g., solids, where there are a less variety of nonreciprocal phenomena in linear response regimes than nonlinear ones due to the need of breaking the time-reversal symmetry. Our finding is distinct from most of the previous studies in open quantum systems that focused on inhomogeneous setups, where a system is coupled to different baths at its boundaries, thus relying on temperature biases or boundary driving. Our open-system formulation is not a response to external electric fields, but allows for featuring direct current generation. In particular, this provides a different framework, for instance, magnetochiral anisotropy, i.e., unidirectional nonlinear resistivity under the magnetic field and electric field for chiral conductors \cite{Rikken01, Morimoto16L, Ideue17}, or transmissions of an electron current in the presence of a potential barrier \cite{Streda03, Birkholz08}. From an experimental perspective, our results can be tested in ultracold atoms; the use of Raman-type spin-orbit coupling is also promissing to break the inversion symmetry. One can also consider semiconductor quantum dots in GaAs as possible experimental candidates \cite{Nitta97, Dirk00, Vandersypen14, Nakajima18}, where the spin relaxation time is very long; spin-resolved dephasing should be realized by using the Zeeman shift.

To summarize, we have proposed minimal setups to realize a nonreciprocal current in open many-body systems. In contrast to conventional approaches in open quantum systems, our finding provides a unique avenue for rectification, namely, the current is neither generated by temperature gradients nor boundary driving, but via the translationally invariant couplings to nonequilibrium baths. We have demonstrated that a nonreciprocal Lindblad operator in general rectifies the current in NESS. We have also revealed that a reciprocal Lindblad operator can be used to rectify the current when the inversion symmetry and the time-reversal symmetry of the internal Hamiltonian are broken. The present analysis opens up various avenues of possible future research such as current rectification in higher dimensions or changes on transport properties by strong integrability breaking.

\begin{acknowledgments}
We are grateful to Takahiro Morimoto and Sota Kitamura for fruitful discussions through the TMS junior researcher visiting program. This work was supported by KAKENHI (Grants No.\ JP18H01140 and No. JP19H01838) and a Grant-in-Aid for Scientific Research on Innovative Areas (KAKENHI Grant No.\ JP15H05855) from the Japan Society for the Promotion of Science. K.Y. was supported by WISE Program, MEXT and JSPS KAKENHI Grant-in-Aid for JSPS fellows Grant No.\ JP20J21318. Y.A. acknowledges support from the Japan Society for the Promotion of Science through Grant No.\ JP19K23424.
\end{acknowledgments}

%%%%%%%[Supplemental Materials]%%%%%%%%%%

%%\clearpage

%%\renewcommand{\thesection}{S\arabic{section}}
%%\renewcommand{\theequation}{S\arabic{equation}}
%%\setcounter{equation}{0}
%%\renewcommand{\thefigure}{S\arabic{figure}}
%%\setcounter{figure}{0}

%%\onecolumngrid
\appendix
%%\begin{center}
%%\large{Supplemental Material for}\\
%%\textbf{"Rectification in Nonequilibrium Steady States of Open Many-Body Systems"}
%%\end{center}

%%%-----[Derivation]-----%%%
\section{Detailed derivation of Lagrange parameters from the time-dependent generalized Gibbs ensemble}
\label{sec_gge}
We explain how the dynamics of the system that is weakly driven by Markovian baths is determined based on the methods considered in Refs.~\cite{Zala17, Zala18a, Zala18b}. We consider a situation that the integrable system described by a Hamiltonian $H_0$ with conservation laws $I_i$ $(i=0, 1, \ldots, N)$ is weakly perturbed by Markovian baths, thus breaking the integrability. Such a system is described by the Lindblad master equation by using a small dimensionless parameter $\epsilon$ as
\begin{gather}
\partial_t \rho =\mathcal L_0\rho  +\mathcal L_1 \rho,\\
\mathcal L_0\rho=-i[H_0, \rho],\quad\mathcal L_1 \rho= \epsilon\mathcal D(\rho),
\end{gather}
where the dissipator $\mathcal D(\rho)$ is given by
\begin{align}
\mathcal D(\rho) =\sum_m\left(L_m\rho L_m^{\dag}-\frac{1}{2}\left\{L_m^{\dag}L_m, \rho\right\}\right).
\end{align}
Below, we take a perturbative approach to the nonequilibrium steady states (NESS). If there is no perturbation by the environments, it is extensively shown that steady states of the integrable models approach to that described by a generalized Gibbs ensemble
\begin{align}
\rho_0 = \frac{e^{-\sum_i \lambda_iI_i}}{\mathrm{Tr}[e^{-\sum_i \lambda_iI_i}]}.
\end{align}
Then, we track the changes of the Lagrange parameters by weak driving of the baths. We split the density operator $\rho(t)$ into zeroth-order approximation $\rho_{\mathrm {GGE}}(t)$ and corrections $\delta \rho(t)$ as
\begin{align}
\rho(t) = \rho_{\mathrm{GGE}}(t) + \delta\rho(t),
\end{align}
where $\rho_{\mathrm{GGE}}(t)$ is the time-dependent generalized Gibbs ensemble (tGGE)
\begin{align}
\rho_{\mathrm{GGE}}(t) = \frac{e^{-\sum_i \lambda_i(t)I_i}}{\mathrm{Tr}[e^{-\sum_i \lambda_i(t)I_i}]}.
\end{align}
and $\delta\rho$ should be small in the limit $\epsilon\to 0$. As $\mathcal L_0\rho_{\mathrm{GGE}}(t)=0$ by definition, the condition of NESS $\mathcal L\rho=0$ ensures that the correction $\delta \rho$ of order of $\epsilon$ (and larger) is given by
\begin{align}
\delta\rho=-\mathcal L^{-1}\mathcal L_1 \rho_{\mathrm{GGE}}.
\end{align}
To obtain the dynamics of Lagrange parameters $\lambda_i(t)$ that determines $\rho_{\mathrm{GGE}}$ of order of $\epsilon^0$, it is convenient to introduce the superoperator $\mathcal P$
\begin{align}
\mathcal P X\equiv - \sum_{ij}\frac{\partial\rho_\mathrm{GGE}}{\partial\lambda_i}(\chi^{-1})_{ij}\mathrm{tr}[I_jX],\label{eq_ProjectionS}\\
\chi_{ij}(t)=\langle I_iI_j\rangle_{\mathrm{GGE}}-\langle I_i\rangle_{\mathrm{GGE}}\langle I_j\rangle_{\mathrm{GGE}},
\end{align}
which projects the density matrix onto the space tangential to the GGE manifold spanned by $\partial \rho_{\mathrm{GGE}}(t)/\partial\lambda_i$. Here, we note that $\mathcal P \rho_{\mathrm{GGE}}\neq\rho_{\mathrm{GGE}}$ because $\mathcal P$ is not a projector onto the space of GGE matrix. By using 
\begin{align}
\mathcal P\dot\rho=\dot\rho_{\mathrm{GGE}}+\mathcal P \delta\dot\rho
\end{align}
and demanding that $\mathcal P \delta\dot\rho\sim O(\epsilon^2)$, we obtain
\begin{align}
\mathcal P \dot \rho\simeq\dot\rho_{\mathrm{GGE}}=\sum_i\frac{\partial\rho_{\mathrm {GGE}}}{\partial\lambda_i}\frac{\partial\lambda_i}{\partial t}
\label{eq_ProjectionS2}
\end{align}
Since $\langle I_i \rangle$ is calculated as
\begin{align}
\langle \dot I_i\rangle &= \mathrm{tr}[I_i\mathcal L \rho]\notag\\
 &= \mathrm{tr}[I_i\mathcal L_1 \rho_{\mathrm{GGE}}]+\mathrm{tr}[I_i\mathcal L _1\delta\rho]\notag\\
 &\simeq\mathrm{tr}[I_i\mathcal L_1 \rho_{\mathrm{GGE}}]
\end{align}
(we have used $\mathrm{tr}[I_i\mathcal L _0\delta\rho]=0$ because $\mathcal L_0^\dag I_i =i[H_0, I_i]=0$, where the adjoint of the Liouvillian is defined by $\mathrm{tr}[A\mathcal L \rho]=\mathrm{tr}[(\mathcal L^\dag A)\rho]$), we finally obtain the dynamics of the Lagrange parameters up to the order of $\epsilon$ from Eqs.~\eqref{eq_ProjectionS} and \eqref{eq_ProjectionS2} as
\begin{align}
\dot \lambda_i =-\sum_j(\chi(t)^{-1})_{ij}\mathrm{tr}\left[I_j\mathcal L_1\rho_{\mathrm{GGE}}(t)\right].
\end{align}
For higher order corrections of the perturbation theory and numerical evidence of the validity of tGGE, see Refs.~\cite{Zala18a, Zala18b}.

\begin{widetext}

\section{Detailed calculations of rate equations for Lagrange parameters}
\label{sec_rate}
We here explain the detailed calculations to obtain the rate equations for Lagrange parameters. For the first model discussed in the main text, the local conservation laws of few-body observables are given by $I_q=c_q^\dag c_q$. Thus, $\chi_{qp}$ in Eq.~\eqref{eq_chi} in the main text is nonzero only for the diagonal components, given by
\begin{align}
 \chi_{qq}(t)
=\langle c_q^\dag c_q c_q^\dag c_q\rangle - \langle c_q^\dag c_q\rangle^2=\langle c_q^\dag c_q\rangle\langle c_q c_q^\dag\rangle=\frac{e^{-\lambda_q}}{(1+e^{-\lambda_q})^2},
 \label{eq_chiS1}
\end{align}
where we have omitted the subscript $\langle\cdots\rangle_{\mathrm{GGE}}$ and the same applies hereafter. Then, we calculate $\langle \dot I_q \rangle=\mathrm{tr}\left[I_q\mathcal L_1 \rho_{\mathrm{GGE}}\right]$ on the right hand side of the rate equation \eqref{eq_rate} for the Lindblad operator \eqref{eq_detailed1} as
\begin{align}
\langle \dot I_q\rangle^{\mathrm{eq}}
&=\frac{\epsilon\gamma_{\mathrm{eq}}}{L}\mathrm{tr}\left[I_q\sum_{kl}\left(L_{kl}^{\mathrm{eq}}\rho_{\mathrm{GGE}} L_{kl}^{\mathrm{eq}\dag}-\frac{1}{2}\left\{L_{kl}^{\mathrm{eq}\dag}L_{kl}^{\mathrm{eq}}, \rho_{\mathrm{GGE}}\right\}\right)\right]\notag\displaybreak[2]\\
&=\frac{\epsilon \gamma_{\mathrm{eq}}J}{L}\mathrm{tr}\left[\sum_{kl}e^{\beta(\epsilon_l-\epsilon_k)/2}c_q^\dag c_q\left(c_k^\dag c_l\rho_{\mathrm{GGE}}c_l^\dag c_k-\frac{1}{2}\left\{c_l^\dag c_kc_k^\dag c_l,\rho_{\mathrm{GGE}}\right\}\right)\right]\notag\displaybreak[2]\\
&=\frac{\epsilon \gamma_{\mathrm{eq}}J}{L}\sum_{kl}e^{\beta(\epsilon_l-\epsilon_k)/2}\left\langle c_l^\dag c_k c_q^\dag c_q c_k^\dag c_l-\frac{1}{2}(c_q^\dag c_q c_l^\dag c_k c_k^\dag c_l + c_l^\dag c_k c_k^\dag c_l c_q^\dag c_q)\right\rangle\notag\displaybreak[2]\\
&=\frac{\epsilon \gamma_{\mathrm{eq}}J}{L}\sum_{k\neq q}\bigg[e^{\beta(\epsilon_k-\epsilon_q)/2}\left(\langle c_k^\dag c_k\rangle\langle c_q c_q^\dag\rangle\langle c_q^\dag c_q\rangle +\langle c_k^\dag c_
k\rangle\langle c_q c_q^\dag\rangle^2\right)\notag\displaybreak[2]\\
&\hspace{2cm}-e^{\beta(\epsilon_q-\epsilon_k)/2}\left(\langle c_q^\dag c_q\rangle^2\langle c_k c_k^\dag\rangle+\langle c_q^\dag c_q\rangle\langle c_q c_q^\dag\rangle\langle c_k c_k^\dag\rangle\right)\bigg]\notag\displaybreak[2]\\
&=\frac{\epsilon \gamma_{\mathrm{eq}}J}{L}\sum_{k}\frac{1}{(1+e^{-\lambda_k})(1+e^{-\lambda_q})}\left(e^{\beta(\epsilon_k-\epsilon_q)/2 -\lambda_k}-e^{\beta(\epsilon_q-\epsilon_k)/2 -\lambda_q}\right),
\label{eq_ForceEqS1}
\end{align}
where we used Wick's theorem. Here, we note that the terms that do not include $q$ in $\sum_{kl}$ becomes zero because such terms correspond to flows $k\to l$ or $l \to k$ $(k, l\neq q)$ and do not contribute to the dynamics of $I_q$. In the same way, we calculate $\langle \dot I_q\rangle$ on the right hand side of the rate equation \eqref{eq_rate} for the Lindblad operator \eqref{eq_dissipation1} as
\begin{align}
\langle \dot I_q\rangle^{\mathrm{neq}}
&=\frac{\epsilon\gamma_{\mathrm{neq}} J}{L}\sum_{q'kk'}\Delta_{kq'}\Delta_{k^\prime q^\prime}^*\left\langle c_{k'}^\dag c_{k'-q'} c_q^\dag c_q c_{k-q'}^\dag c_k-\frac{1}{2}(c_q^\dag c_q c_{k'}^\dag c_{k'-q'} c_{k-q'}^\dag c_k + c_{k'}^\dag c_{k'-q'} c_{k-q'}^\dag c_k c_q^\dag c_q)\right\rangle\notag\\
&=\frac{\epsilon \gamma_{\mathrm{neq}} J}{L}\sum_k\left(-|\Delta_{qk}|^2\langle c_q^\dag c_q\rangle \langle c_{q-k} c_{q-k}^\dag\rangle+|\Delta_{q+k,k}|^2\langle c_{q+k}^\dag c_{q+k}\rangle \langle c_q c_q^\dag\rangle\right)\notag\\
&=\frac{\epsilon \gamma_{\mathrm{neq}} J}{L}\sum_k\Bigg[-|\Delta_{q,q-k}|^2\frac{e^{-\lambda_q}}{(1+e^{-\lambda_k})(1+e^{-\lambda_q})}+|\Delta_{k, k-q}|^2\frac{e^{-\lambda_k}}{(1+e^{-\lambda_{k}})(1+e^{-\lambda_q})}\Bigg].
\label{eq_ForceNeqS1}
\end{align}
By using Eqs.~\eqref{eq_chiS1}--\eqref{eq_ForceNeqS1}, we obtain the rate equation \eqref{eq_rate1} for the first model in the main text.

For the second model (with the reciprocal dissipator), we can calculate the rate equation almost in the same way as discussed above. As the local conservation law is given by $I_{q\nu}=\eta_{q\nu}^\dag\eta_{q\nu}$ ($\nu=\pm$), $\chi_{q\nu, p\mu}$ in Eq.~\eqref{eq_chi} in the main text is zero for the off-diagonal components and the diagonal component is calculated as 
\begin{align}
\chi_{q\nu, q\nu}(t)
=\langle \eta_{q\nu}^\dag \eta_{q\nu} \eta_{q\nu}^\dag \eta_{q\nu}\rangle - \left[\langle \eta_{q\nu}^\dag \eta_{q\nu}\rangle\right]^2
=\langle \eta_{q\nu}^\dag \eta_{q\nu}\rangle\langle \eta_{q\nu} \eta_{q\nu}^\dag\rangle
=\frac{e^{-\lambda_{q\nu}}}{(1+e^{-\lambda_{q\nu}})^2}.
\label{eq_chiS2}
\end{align}
We see from Eq.~\eqref{eq_chiS2} that the degrees of freedom in momentum space are doubled by upper and lower energy bands compared to Eq.~\eqref{eq_chiS1}. Then, $\langle \dot I_{q\nu}\rangle$ on the right hand side of the rate equation \eqref{eq_rate} for the Lindblad operator \eqref{eq_detailed2} is calculated by doubling the momentum space as [see also Eq.~\eqref{eq_ForceEqS1}]
\begin{align}
\langle\dot I_{q\nu}\rangle^{\mathrm{eq}}
&=\frac{\epsilon\gamma_{\mathrm{eq}}J}{L}\sum_{kl}\sum_{\mu,\kappa=\pm}e^{\beta(\epsilon_{l\kappa}-\epsilon_{k\mu})/2}\left\langle \eta_{l\kappa}^\dag \eta_{k\mu} \eta_{q\nu}^\dag \eta_{q\nu} \eta_{k\mu}^\dag \eta_{l\kappa}-\frac{1}{2}(\eta_{q\nu}^\dag \eta_{q\nu} \eta_{l\kappa}^\dag \eta_{k\mu} \eta_{k\mu}^\dag \eta_{l\kappa} + \eta_{l\kappa}^\dag \eta_{k\mu} \eta_{k\mu}^\dag \eta_{l\kappa} \eta_{q\nu}^\dag \eta_{q\nu})\right\rangle\notag\\
&=\frac{\epsilon\gamma_{\mathrm{eq}}J}{L}\sum_{k\mu}\frac{1}{(1+e^{-\lambda_{k\mu}})(1+e^{-\lambda_{q\nu}})}\left(e^{\beta(\epsilon_{k\mu}-\epsilon_{q\nu})/2 -\lambda_{k\mu}}-e^{\beta(\epsilon_{q\nu}-\epsilon_{k\mu})/2 -\lambda_{q\nu}}\right).
\end{align}
The contribution from the nonequilibrium bath, denoted as $\langle \dot I_{q\nu}\rangle^{\mathrm{neq}}$, can also be simplified by using the expression of the Lindblad operator \eqref{eq_dissipation2}:
\begin{align}
\langle\dot I_{q\nu}\rangle^{\mathrm{neq}}=\frac{\epsilon\gamma_{\mathrm{neq}} J}{L}\sum_{k k^\prime q^\prime, \sigma=\uparrow\downarrow}\gamma_\sigma\Bigg\langle c_{k\sigma}^\dag c_{k-q',\sigma}\eta_{q\nu}^\dag\eta_{q\nu} c_{k^\prime-q', \sigma}^\dag c_{k^\prime\sigma}&-\frac{1}{2}\eta_{q\nu}^\dag\eta_{q\nu} c_{k\sigma}^\dag c_{k-q',\sigma} c_{k^\prime-q',\sigma}^\dag c_{k^\prime\sigma}\notag\\
&-\frac{1}{2} c_{k\sigma}^\dag c_{k-q',\sigma} c_{k^\prime-q', \sigma}^\dag c_{k^\prime\sigma}\eta_{q\nu}^\dag\eta_{q\nu}\Bigg\rangle.
\label{eq_ForceNeqS2a}
\end{align}
To use Wick's theorem, we substitute the Bogoliubov transformation $c_{k\sigma}=u_{\sigma \nu}(k)\eta_{k \nu}$ in Eq.~\eqref{eq_ForceNeqS2a} (for the detailed form of $u_{\sigma\nu}(k)$, see Appendix~\ref{sec_quasi}). We note that, though we have to calculate $2^4$ times as many terms as Eq.~\eqref{eq_ForceNeqS2a} as a result of the substitution, many of which become zero since tGGE ensemble is defined by local conservation quantities. Then, we obtain
\begin{align}
\langle\dot I_{q\nu}\rangle^{\mathrm{neq}}
&=\frac{\epsilon\gamma_{\mathrm{neq}} J}{L}\sum_k\sum_{\mu=\pm}\sum_{\sigma=\uparrow\downarrow}\gamma_\sigma|u_{\sigma\nu}(q)|^2|u_{\sigma\mu}(k)|^2\frac{e^{-\lambda_{k\mu}}-e^{-\lambda_{q\nu}}}{(1+e^{-\lambda_{k\mu}})(1+e^{-\lambda_{q\nu}})}.
\label{eq_ForceNeqS2b}
\end{align}
Finally, Eq.~\eqref{eq_rate2} in the main text follows from Eqs.~\eqref{eq_chiS2}--\eqref{eq_ForceNeqS2b}.

\section{Detailed derivation of the quasiparticle operators}
\label{sec_quasi}
Here, we explain the detailed derivation of the quasiparticle operators for the second model in the main text. The tight-binding Hamiltonian with the Rashba spin-orbit coupling and the Zeeman magnetic field (Eq.~\eqref{eq_Rashba} in the main text) is diagonalized as 
\begin{align}
H_0 = &-J\sum_{j\sigma}(c_{j+1\sigma}^\dag c_{j\sigma} + \mathrm{H.c.}) + h\sum_j(n_{j\uparrow}-n_{j\downarrow}) \notag\displaybreak[2]\\
&-\alpha_z\sum_{j\sigma\sigma^\prime}(c_{j+1\sigma}^\dag(i\sigma_y)_{\sigma\sigma^\prime} c_{j\sigma^\prime} + \mathrm{H.c.}) + \alpha_y\sum_{j\sigma\sigma^\prime}(c_{j+1\sigma}^\dag(i\sigma_z)_{\sigma\sigma^\prime} c_{j\sigma^\prime} + \mathrm{H.c.})\notag\displaybreak[2]\\
=&\sum_k
\left(
\begin{matrix}
c_{k\uparrow}^\dag&c_{k\downarrow}^\dag
\end{matrix}
\right)
\left(
\begin{matrix}
-2J \cos k+2\alpha_y\sin k+h&2i\alpha_z\sin k\\
-2i\alpha_z\sin k&-2J \cos k -2\alpha_y\sin k -h
\end{matrix}
\right)
\left(
\begin{matrix}
c_{k\uparrow}\\
c_{k\downarrow}
\end{matrix}
\right)\notag\\
=&\sum_{k,\nu=\pm}\epsilon_{k\nu}\eta_{k\nu}^\dag\eta_{k\nu},
\end{align}
with eigenvalues
\begin{align}
\epsilon_{k\pm}=-2J\cos (k)\pm\sqrt{(2\alpha_y \sin (k)+h)^2 +4\alpha_z^2\sin^2 (k)},
\end{align}
and quasiparticles, which are given by the unitary transformation,
\begin{align}
\left(
\begin{matrix}
c_{k\uparrow}\\
c_{k\downarrow}
\end{matrix}
\right)
=U(k)\left(
\begin{matrix}
\eta_{k+}\\
\eta_{k-}
\end{matrix}
\right),
\end{align}
\begin{align}
U(k)=
\left(
\begin{matrix}
u_{\uparrow +}(k)& u_{\uparrow -}(k)\\
u_{\downarrow +}(k)& u_{\downarrow -}(k)
\end{matrix}
\right)
=\frac{1}{\sqrt 2}
\left(
\begin{matrix}
-i\sqrt{\frac{2\alpha_y \sin k+h}{\sqrt{(2\alpha_y \sin k+h)^2 +4\alpha_z^2\sin^2 k}}+1}
&-i\sqrt{\frac{-2\alpha_y \sin k-h}{\sqrt{(2\alpha_y \sin k+h)^2 +4\alpha_z^2\sin^2 k}}+1}\\
-\frac{\sin(k)}{|\sin(k)|}\sqrt{\frac{-2\alpha_y \sin k-h}{\sqrt{(2\alpha_y \sin k+h)^2 +4\alpha_z^2\sin^2 k}}+1}
&\frac{\sin(k)}{|\sin(k)|}\sqrt{\frac{2\alpha_y \sin k+h}{\sqrt{(2\alpha_y \sin k+h)^2 +4\alpha_z^2\sin^2 k}}+1}
\end{matrix}
\right).
\label{eq_unitaryS}
\end{align}
We see from Eq.~\eqref{eq_Fneq2} in the main text and Eq.~\eqref{eq_unitaryS} that the contribution to Lagrange parameters from the nonequilibrium bath is inversion symmetric with respect to $q$ if either one of  the Zeeman magnetic field or the Rashba spin-orbit coupling is absent, which does not rectify the current. As a result, we need to break both the inversion symmetry and the time-reversal symmetry of the Hamiltonian to obtain the nonreciprocal current in NESS. We can also calculate the current \eqref{eq_current2} in the main text by using these quasiparticle operators as
\begin{align}
I_\uparrow
=& \frac{2J}{L}\sum_q\sin (q) \langle c_{q\uparrow}^\dag c_{q\uparrow}\rangle +\frac{2\alpha_y}{L}\sum_q\cos(q)\langle c_{q\uparrow}^\dag c_{q\uparrow}\rangle + \frac{2i\alpha_z}{L} \sum_q \cos (q)\langle c_{q\uparrow}^\dag c_{q\downarrow}\rangle\notag\displaybreak[2]\\
=&\frac{1}{L}\sum_q \left(J\sin(q)+\alpha_y\cos(q)\right)\Bigg\langle \left(\frac{2\alpha_y\sin(q) + h}{\sqrt{(2\alpha_y \sin q+h)^2 +4\alpha_z^2\sin^2 q}}+1\right)\eta_{q+}^\dag\eta_{q+}\notag\displaybreak[2]\\
&\hspace{6cm}+ \left(\frac{-2\alpha_y\sin(q) - h}{\sqrt{(2\alpha_y \sin q+h)^2 +4\alpha_z^2\sin^2 q}}+1\right)\eta_{q-}^\dag\eta_{q-}\Bigg\rangle\notag\\
&+\frac{2\alpha_z^2}{L}\sum_q\cos(q)\sin(q)\left\langle\frac{\eta_{q+}^\dag\eta_{q+}-\eta_{q-}^\dag\eta_{q-}}{\sqrt{(2\alpha_y \sin q+h)^2 +4\alpha_z^2\sin^2 q}}\right\rangle
\label{eq_currentS1}
\end{align}
\begin{align}
I_\downarrow
=& \frac{2J}{L}\sum_q\sin (q) \langle c_{q\downarrow}^\dag c_{q\downarrow}\rangle -\frac{2\alpha_y}{L}\sum_q\cos(q)\langle c_{q\downarrow}^\dag c_{q\downarrow}\rangle - \frac{2i\alpha_z}{L} \sum_q \cos (q)\langle c_{q\downarrow}^\dag c_{q\uparrow}\notag\displaybreak[2]\\
=&\frac{1}{L}\sum_q \left(J\sin(q)-\alpha_y\cos(q)\right)\Bigg\langle \left(\frac{-2\alpha_y\sin(q) - h}{\sqrt{(2\alpha_y \sin k+h)^2 +4\alpha_z^2\sin^2 k}}+1\right)\eta_{k+}^\dag\eta_{k+}\notag\displaybreak[2]\\
&\hspace{6cm}+ \left(\frac{2\alpha_y\sin(q) + h}{\sqrt{(2\alpha_y \sin q+h)^2 +4\alpha_z^2\sin^2 q}}+1\right)\eta_{q-}^\dag\eta_{q-}\Bigg\rangle\notag\displaybreak[2]\\
&+\frac{2\alpha_z^2}{L}\sum_q\cos(q)\sin(q)\left\langle\frac{\eta_{q+}^\dag\eta_{q+}-\eta_{q-}^\dag\eta_{q-}}{\sqrt{(2\alpha_y \sin q+h)^2 +4\alpha_z^2\sin^2 q}}\right\rangle.
\label{eq_currentS2}
\end{align}

\section{Results of the nonreciprocal current in NESS with down-spin dephasing}
\label{sec_downspin}
We here give the numerical results of the current in NESS when dephasing is applied to down spins in the second model discussed in the main text. From Fig.~\ref{fig6}(a), we see that the current rectifies in the opposite direction and the total current $I$ has the reversed value of that in Fig.~\ref{fig4}(a) in the main text. As shown in Fig.~\ref{fig6}(b), the change of population near the Fermi surface where dephasing is applied becomes large (grey dotted circles) compared to the Gibbs state, which contributes to the current in the negative direction [see Eqs.~\eqref{eq_currentS1} and \eqref{eq_currentS2}].
\begin{figure}[b]
\includegraphics[width=15cm]{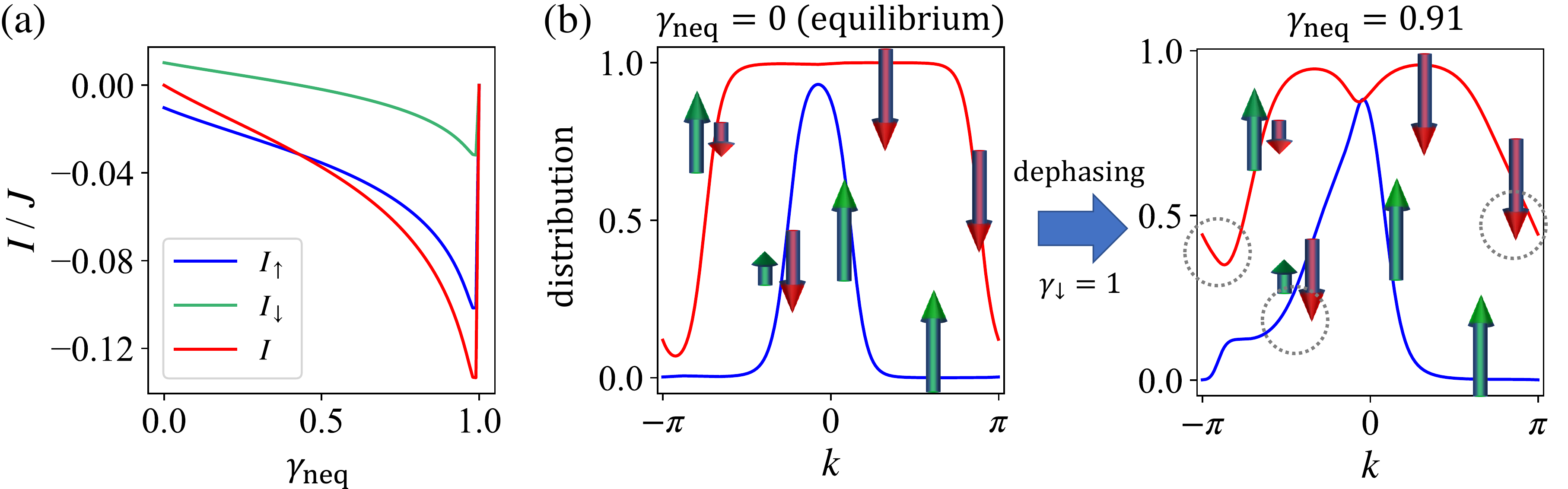}
\caption{(a) Current in NESS and its spin dependence as a function of $\gamma_{\mathrm{neq}}$ in the presence of the Zeeman magnetic field and the Rashba spin-orbit coupling (model 2 in the main text), where dephasing is applied to down spins. (b) Distribution of the upper band (blue) and the lower band (red) in NESS for the equilibrium Gibbs state (left) and the nonequilibrium state (right) corresponding to (a). Change of population near the Fermi surface where dephasing is applied becomes large (marked by grey dotted circles). The parameters are set to $\beta=2/J$, $\alpha_y=1.1J$, $\alpha_z=0.9J$, and $h=J$ for the initial state at infinite temperature.}
\label{fig6}
\end{figure}

\end{widetext}

\bibliographystyle{apsrev4-1}
\bibliography{Rectification.bib}

\providecommand{\noopsort}[1]{}\providecommand{\singleletter}[1]{#1}%
\begin{thebibliography}{110}%
\makeatletter
\providecommand \@ifxundefined [1]{%
 \@ifx{#1\undefined}
}%
\providecommand \@ifnum [1]{%
 \ifnum #1\expandafter \@firstoftwo
 \else \expandafter \@secondoftwo
 \fi
}%
\providecommand \@ifx [1]{%
 \ifx #1\expandafter \@firstoftwo
 \else \expandafter \@secondoftwo
 \fi
}%
\providecommand \natexlab [1]{#1}%
\providecommand \enquote  [1]{``#1''}%
\providecommand \bibnamefont  [1]{#1}%
\providecommand \bibfnamefont [1]{#1}%
\providecommand \citenamefont [1]{#1}%
\providecommand \href@noop [0]{\@secondoftwo}%
\providecommand \href [0]{\begingroup \@sanitize@url \@href}%
\providecommand \@href[1]{\@@startlink{#1}\@@href}%
\providecommand \@@href[1]{\endgroup#1\@@endlink}%
\providecommand \@sanitize@url [0]{\catcode `\\12\catcode `\$12\catcode
  `\&12\catcode `\#12\catcode `\^12\catcode `\_12\catcode `\%12\relax}%
\providecommand \@@startlink[1]{}%
\providecommand \@@endlink[0]{}%
\providecommand \url  [0]{\begingroup\@sanitize@url \@url }%
\providecommand \@url [1]{\endgroup\@href {#1}{\urlprefix }}%
\providecommand \urlprefix  [0]{URL }%
\providecommand \Eprint [0]{\href }%
\providecommand \doibase [0]{http://dx.doi.org/}%
\providecommand \selectlanguage [0]{\@gobble}%
\providecommand \bibinfo  [0]{\@secondoftwo}%
\providecommand \bibfield  [0]{\@secondoftwo}%
\providecommand \translation [1]{[#1]}%
\providecommand \BibitemOpen [0]{}%
\providecommand \bibitemStop [0]{}%
\providecommand \bibitemNoStop [0]{.\EOS\space}%
\providecommand \EOS [0]{\spacefactor3000\relax}%
\providecommand \BibitemShut  [1]{\csname bibitem#1\endcsname}%
\let\auto@bib@innerbib\@empty
%</preamble>
\bibitem [{\citenamefont {Diehl}\ \emph {et~al.}(2008)\citenamefont {Diehl},
  \citenamefont {Micheli}, \citenamefont {Kantian}, \citenamefont {Kraus},
  \citenamefont {B{\"u}chler},\ and\ \citenamefont {Zoller}}]{Diehl08}%
  \BibitemOpen
  \bibfield  {author} {\bibinfo {author} {\bibfnamefont {S.}~\bibnamefont
  {Diehl}}, \bibinfo {author} {\bibfnamefont {A.}~\bibnamefont {Micheli}},
  \bibinfo {author} {\bibfnamefont {A.}~\bibnamefont {Kantian}}, \bibinfo
  {author} {\bibfnamefont {B.}~\bibnamefont {Kraus}}, \bibinfo {author}
  {\bibfnamefont {H.}~\bibnamefont {B{\"u}chler}}, \ and\ \bibinfo {author}
  {\bibfnamefont {P.}~\bibnamefont {Zoller}},\ }\href@noop {} {\bibfield
  {journal} {\bibinfo  {journal} {Nat. Phys.}\ }\textbf {\bibinfo {volume}
  {4}},\ \bibinfo {pages} {878} (\bibinfo {year} {2008})}\BibitemShut {NoStop}%
\bibitem [{\citenamefont {Kraus}\ \emph {et~al.}(2008)\citenamefont {Kraus},
  \citenamefont {B\"uchler}, \citenamefont {Diehl}, \citenamefont {Kantian},
  \citenamefont {Micheli},\ and\ \citenamefont {Zoller}}]{Kraus08}%
  \BibitemOpen
  \bibfield  {author} {\bibinfo {author} {\bibfnamefont {B.}~\bibnamefont
  {Kraus}}, \bibinfo {author} {\bibfnamefont {H.~P.}\ \bibnamefont
  {B\"uchler}}, \bibinfo {author} {\bibfnamefont {S.}~\bibnamefont {Diehl}},
  \bibinfo {author} {\bibfnamefont {A.}~\bibnamefont {Kantian}}, \bibinfo
  {author} {\bibfnamefont {A.}~\bibnamefont {Micheli}}, \ and\ \bibinfo
  {author} {\bibfnamefont {P.}~\bibnamefont {Zoller}},\ }\href {\doibase
  10.1103/PhysRevA.78.042307} {\bibfield  {journal} {\bibinfo  {journal} {Phys.
  Rev. A}\ }\textbf {\bibinfo {volume} {78}},\ \bibinfo {pages} {042307}
  (\bibinfo {year} {2008})}\BibitemShut {NoStop}%
\bibitem [{\citenamefont {M{\"u}ller}\ \emph {et~al.}(2012)\citenamefont
  {M{\"u}ller}, \citenamefont {Diehl}, \citenamefont {Pupillo},\ and\
  \citenamefont {Zoller}}]{Muller12}%
  \BibitemOpen
  \bibfield  {author} {\bibinfo {author} {\bibfnamefont {M.}~\bibnamefont
  {M{\"u}ller}}, \bibinfo {author} {\bibfnamefont {S.}~\bibnamefont {Diehl}},
  \bibinfo {author} {\bibfnamefont {G.}~\bibnamefont {Pupillo}}, \ and\
  \bibinfo {author} {\bibfnamefont {P.}~\bibnamefont {Zoller}},\ }\href@noop {}
  {\bibfield  {journal} {\bibinfo  {journal} {Adv. Atom. Mol. Opt. Phys.}\
  }\textbf {\bibinfo {volume} {61}},\ \bibinfo {pages} {1} (\bibinfo {year}
  {2012})}\BibitemShut {NoStop}%
\bibitem [{\citenamefont {Daley}(2014)}]{Daley14}%
  \BibitemOpen
  \bibfield  {author} {\bibinfo {author} {\bibfnamefont {A.~J.}\ \bibnamefont
  {Daley}},\ }\href@noop {} {\bibfield  {journal} {\bibinfo  {journal} {Adv.
  Phys.}\ }\textbf {\bibinfo {volume} {63}},\ \bibinfo {pages} {77} (\bibinfo
  {year} {2014})}\BibitemShut {NoStop}%
\bibitem [{\citenamefont {Ikeda}\ and\ \citenamefont {Sato}(2020)}]{Ikeda20}%
  \BibitemOpen
  \bibfield  {author} {\bibinfo {author} {\bibfnamefont {T.~N.}\ \bibnamefont
  {Ikeda}}\ and\ \bibinfo {author} {\bibfnamefont {M.}~\bibnamefont {Sato}},\
  }\href@noop {} {\bibfield  {journal} {\bibinfo  {journal} {Sci. Adv.}\
  }\textbf {\bibinfo {volume} {6}},\ \bibinfo {pages} {eabb4019} (\bibinfo
  {year} {2020})}\BibitemShut {NoStop}%
\bibitem [{\citenamefont {Ashida}\ \emph {et~al.}()\citenamefont {Ashida},
  \citenamefont {Gong},\ and\ \citenamefont {Ueda}}]{Ashida20}%
  \BibitemOpen
  \bibfield  {author} {\bibinfo {author} {\bibfnamefont {Y.}~\bibnamefont
  {Ashida}}, \bibinfo {author} {\bibfnamefont {Z.}~\bibnamefont {Gong}}, \ and\
  \bibinfo {author} {\bibfnamefont {M.}~\bibnamefont {Ueda}},\ }\href@noop {}
  {\bibinfo  {journal} {arXiv:2006.01837}\ }\BibitemShut {NoStop}%
\bibitem [{\citenamefont {Sieberer}\ \emph {et~al.}(2013)\citenamefont
  {Sieberer}, \citenamefont {Huber}, \citenamefont {Altman},\ and\
  \citenamefont {Diehl}}]{Sieberer13}%
  \BibitemOpen
\bibfield  {journal} {  }\bibfield  {author} {\bibinfo {author} {\bibfnamefont
  {L.~M.}\ \bibnamefont {Sieberer}}, \bibinfo {author} {\bibfnamefont {S.~D.}\
  \bibnamefont {Huber}}, \bibinfo {author} {\bibfnamefont {E.}~\bibnamefont
  {Altman}}, \ and\ \bibinfo {author} {\bibfnamefont {S.}~\bibnamefont
  {Diehl}},\ }\href {\doibase 10.1103/PhysRevLett.110.195301} {\bibfield
  {journal} {\bibinfo  {journal} {Phys. Rev. Lett.}\ }\textbf {\bibinfo
  {volume} {110}},\ \bibinfo {pages} {195301} (\bibinfo {year}
  {2013})}\BibitemShut {NoStop}%
\bibitem [{\citenamefont {Ashida}\ \emph {et~al.}(2017)\citenamefont {Ashida},
  \citenamefont {Furukawa},\ and\ \citenamefont {Ueda}}]{Ashida17}%
  \BibitemOpen
  \bibfield  {author} {\bibinfo {author} {\bibfnamefont {Y.}~\bibnamefont
  {Ashida}}, \bibinfo {author} {\bibfnamefont {S.}~\bibnamefont {Furukawa}}, \
  and\ \bibinfo {author} {\bibfnamefont {M.}~\bibnamefont {Ueda}},\ }\href
  {\doibase 10.1038/ncomms15791} {\bibfield  {journal} {\bibinfo  {journal}
  {Nat. Commun.}\ }\textbf {\bibinfo {volume} {8}},\ \bibinfo {pages} {15791}
  (\bibinfo {year} {2017})}\BibitemShut {NoStop}%
\bibitem [{\citenamefont {Ashida}\ \emph {et~al.}(2016)\citenamefont {Ashida},
  \citenamefont {Furukawa},\ and\ \citenamefont {Ueda}}]{Ashida16}%
  \BibitemOpen
  \bibfield  {author} {\bibinfo {author} {\bibfnamefont {Y.}~\bibnamefont
  {Ashida}}, \bibinfo {author} {\bibfnamefont {S.}~\bibnamefont {Furukawa}}, \
  and\ \bibinfo {author} {\bibfnamefont {M.}~\bibnamefont {Ueda}},\ }\href
  {\doibase 10.1103/PhysRevA.94.053615} {\bibfield  {journal} {\bibinfo
  {journal} {Phys. Rev. A}\ }\textbf {\bibinfo {volume} {94}},\ \bibinfo
  {pages} {053615} (\bibinfo {year} {2016})}\BibitemShut {NoStop}%
\bibitem [{\citenamefont {Nakagawa}\ \emph {et~al.}(2018)\citenamefont
  {Nakagawa}, \citenamefont {Kawakami},\ and\ \citenamefont
  {Ueda}}]{Nakagawa18}%
  \BibitemOpen
  \bibfield  {author} {\bibinfo {author} {\bibfnamefont {M.}~\bibnamefont
  {Nakagawa}}, \bibinfo {author} {\bibfnamefont {N.}~\bibnamefont {Kawakami}},
  \ and\ \bibinfo {author} {\bibfnamefont {M.}~\bibnamefont {Ueda}},\ }\href
  {\doibase 10.1103/PhysRevLett.121.203001} {\bibfield  {journal} {\bibinfo
  {journal} {Phys. Rev. Lett.}\ }\textbf {\bibinfo {volume} {121}},\ \bibinfo
  {pages} {203001} (\bibinfo {year} {2018})}\BibitemShut {NoStop}%
\bibitem [{\citenamefont {Diehl}\ \emph
  {et~al.}(2010{\natexlab{a}})\citenamefont {Diehl}, \citenamefont {Tomadin},
  \citenamefont {Micheli}, \citenamefont {Fazio},\ and\ \citenamefont
  {Zoller}}]{Diehl10}%
  \BibitemOpen
  \bibfield  {author} {\bibinfo {author} {\bibfnamefont {S.}~\bibnamefont
  {Diehl}}, \bibinfo {author} {\bibfnamefont {A.}~\bibnamefont {Tomadin}},
  \bibinfo {author} {\bibfnamefont {A.}~\bibnamefont {Micheli}}, \bibinfo
  {author} {\bibfnamefont {R.}~\bibnamefont {Fazio}}, \ and\ \bibinfo {author}
  {\bibfnamefont {P.}~\bibnamefont {Zoller}},\ }\href {\doibase
  10.1103/PhysRevLett.105.015702} {\bibfield  {journal} {\bibinfo  {journal}
  {Phys. Rev. Lett.}\ }\textbf {\bibinfo {volume} {105}},\ \bibinfo {pages}
  {015702} (\bibinfo {year} {2010}{\natexlab{a}})}\BibitemShut {NoStop}%
\bibitem [{\citenamefont {H\"oning}\ \emph {et~al.}(2012)\citenamefont
  {H\"oning}, \citenamefont {Moos},\ and\ \citenamefont
  {Fleischhauer}}]{Honing12}%
  \BibitemOpen
  \bibfield  {author} {\bibinfo {author} {\bibfnamefont {M.}~\bibnamefont
  {H\"oning}}, \bibinfo {author} {\bibfnamefont {M.}~\bibnamefont {Moos}}, \
  and\ \bibinfo {author} {\bibfnamefont {M.}~\bibnamefont {Fleischhauer}},\
  }\href {\doibase 10.1103/PhysRevA.86.013606} {\bibfield  {journal} {\bibinfo
  {journal} {Phys. Rev. A}\ }\textbf {\bibinfo {volume} {86}},\ \bibinfo
  {pages} {013606} (\bibinfo {year} {2012})}\BibitemShut {NoStop}%
\bibitem [{\citenamefont {Hamazaki}\ \emph {et~al.}(2019)\citenamefont
  {Hamazaki}, \citenamefont {Kawabata},\ and\ \citenamefont
  {Ueda}}]{Hamazaki19}%
  \BibitemOpen
  \bibfield  {author} {\bibinfo {author} {\bibfnamefont {R.}~\bibnamefont
  {Hamazaki}}, \bibinfo {author} {\bibfnamefont {K.}~\bibnamefont {Kawabata}},
  \ and\ \bibinfo {author} {\bibfnamefont {M.}~\bibnamefont {Ueda}},\ }\href
  {\doibase 10.1103/PhysRevLett.123.090603} {\bibfield  {journal} {\bibinfo
  {journal} {Phys. Rev. Lett.}\ }\textbf {\bibinfo {volume} {123}},\ \bibinfo
  {pages} {090603} (\bibinfo {year} {2019})}\BibitemShut {NoStop}%
\bibitem [{\citenamefont {Yamamoto}\ \emph {et~al.}(2019)\citenamefont
  {Yamamoto}, \citenamefont {Nakagawa}, \citenamefont {Adachi}, \citenamefont
  {Takasan}, \citenamefont {Ueda},\ and\ \citenamefont
  {Kawakami}}]{Yamamoto19}%
  \BibitemOpen
  \bibfield  {author} {\bibinfo {author} {\bibfnamefont {K.}~\bibnamefont
  {Yamamoto}}, \bibinfo {author} {\bibfnamefont {M.}~\bibnamefont {Nakagawa}},
  \bibinfo {author} {\bibfnamefont {K.}~\bibnamefont {Adachi}}, \bibinfo
  {author} {\bibfnamefont {K.}~\bibnamefont {Takasan}}, \bibinfo {author}
  {\bibfnamefont {M.}~\bibnamefont {Ueda}}, \ and\ \bibinfo {author}
  {\bibfnamefont {N.}~\bibnamefont {Kawakami}},\ }\href {\doibase
  10.1103/PhysRevLett.123.123601} {\bibfield  {journal} {\bibinfo  {journal}
  {Phys. Rev. Lett.}\ }\textbf {\bibinfo {volume} {123}},\ \bibinfo {pages}
  {123601} (\bibinfo {year} {2019})}\BibitemShut {NoStop}%
\bibitem [{\citenamefont {Matsumoto}\ \emph {et~al.}()\citenamefont
  {Matsumoto}, \citenamefont {Kawabata}, \citenamefont {Ashida}, \citenamefont
  {Furukawa},\ and\ \citenamefont {Ueda}}]{Matsumoto19}%
  \BibitemOpen
  \bibfield  {author} {\bibinfo {author} {\bibfnamefont {N.}~\bibnamefont
  {Matsumoto}}, \bibinfo {author} {\bibfnamefont {K.}~\bibnamefont {Kawabata}},
  \bibinfo {author} {\bibfnamefont {Y.}~\bibnamefont {Ashida}}, \bibinfo
  {author} {\bibfnamefont {S.}~\bibnamefont {Furukawa}}, \ and\ \bibinfo
  {author} {\bibfnamefont {M.}~\bibnamefont {Ueda}},\ }\href@noop {} {\bibinfo
  {journal} {arXiv:1912.09045}\ }\BibitemShut {NoStop}%
\bibitem [{\citenamefont {Diehl}\ \emph
  {et~al.}(2010{\natexlab{b}})\citenamefont {Diehl}, \citenamefont {Yi},
  \citenamefont {Daley},\ and\ \citenamefont {Zoller}}]{Diehl10a}%
  \BibitemOpen
\bibfield  {journal} {  }\bibfield  {author} {\bibinfo {author} {\bibfnamefont
  {S.}~\bibnamefont {Diehl}}, \bibinfo {author} {\bibfnamefont
  {W.}~\bibnamefont {Yi}}, \bibinfo {author} {\bibfnamefont {A.~J.}\
  \bibnamefont {Daley}}, \ and\ \bibinfo {author} {\bibfnamefont
  {P.}~\bibnamefont {Zoller}},\ }\href {\doibase
  10.1103/PhysRevLett.105.227001} {\bibfield  {journal} {\bibinfo  {journal}
  {Phys. Rev. Lett.}\ }\textbf {\bibinfo {volume} {105}},\ \bibinfo {pages}
  {227001} (\bibinfo {year} {2010}{\natexlab{b}})}\BibitemShut {NoStop}%
\bibitem [{\citenamefont {Yi}\ \emph {et~al.}(2012)\citenamefont {Yi},
  \citenamefont {Diehl}, \citenamefont {Daley},\ and\ \citenamefont
  {Zoller}}]{Yi12}%
  \BibitemOpen
  \bibfield  {author} {\bibinfo {author} {\bibfnamefont {W.}~\bibnamefont
  {Yi}}, \bibinfo {author} {\bibfnamefont {S.}~\bibnamefont {Diehl}}, \bibinfo
  {author} {\bibfnamefont {A.}~\bibnamefont {Daley}}, \ and\ \bibinfo {author}
  {\bibfnamefont {P.}~\bibnamefont {Zoller}},\ }\href@noop {} {\bibfield
  {journal} {\bibinfo  {journal} {New J. Phys.}\ }\textbf {\bibinfo {volume}
  {14}},\ \bibinfo {pages} {055002} (\bibinfo {year} {2012})}\BibitemShut
  {NoStop}%
\bibitem [{\citenamefont {Nakagawa}\ \emph {et~al.}(2020)\citenamefont
  {Nakagawa}, \citenamefont {Tsuji}, \citenamefont {Kawakami},\ and\
  \citenamefont {Ueda}}]{Nakagawa20}%
  \BibitemOpen
  \bibfield  {author} {\bibinfo {author} {\bibfnamefont {M.}~\bibnamefont
  {Nakagawa}}, \bibinfo {author} {\bibfnamefont {N.}~\bibnamefont {Tsuji}},
  \bibinfo {author} {\bibfnamefont {N.}~\bibnamefont {Kawakami}}, \ and\
  \bibinfo {author} {\bibfnamefont {M.}~\bibnamefont {Ueda}},\ }\href {\doibase
  10.1103/PhysRevLett.124.147203} {\bibfield  {journal} {\bibinfo  {journal}
  {Phys. Rev. Lett.}\ }\textbf {\bibinfo {volume} {124}},\ \bibinfo {pages}
  {147203} (\bibinfo {year} {2020})}\BibitemShut {NoStop}%
\bibitem [{\citenamefont {D\"urr}\ \emph {et~al.}(2009)\citenamefont {D\"urr},
  \citenamefont {Garc\'{\i}a-Ripoll}, \citenamefont {Syassen}, \citenamefont
  {Bauer}, \citenamefont {Lettner}, \citenamefont {Cirac},\ and\ \citenamefont
  {Rempe}}]{Durr09}%
  \BibitemOpen
  \bibfield  {author} {\bibinfo {author} {\bibfnamefont {S.}~\bibnamefont
  {D\"urr}}, \bibinfo {author} {\bibfnamefont {J.~J.}\ \bibnamefont
  {Garc\'{\i}a-Ripoll}}, \bibinfo {author} {\bibfnamefont {N.}~\bibnamefont
  {Syassen}}, \bibinfo {author} {\bibfnamefont {D.~M.}\ \bibnamefont {Bauer}},
  \bibinfo {author} {\bibfnamefont {M.}~\bibnamefont {Lettner}}, \bibinfo
  {author} {\bibfnamefont {J.~I.}\ \bibnamefont {Cirac}}, \ and\ \bibinfo
  {author} {\bibfnamefont {G.}~\bibnamefont {Rempe}},\ }\href {\doibase
  10.1103/PhysRevA.79.023614} {\bibfield  {journal} {\bibinfo  {journal} {Phys.
  Rev. A}\ }\textbf {\bibinfo {volume} {79}},\ \bibinfo {pages} {023614}
  (\bibinfo {year} {2009})}\BibitemShut {NoStop}%
\bibitem [{\citenamefont {Ashida}\ and\ \citenamefont {Ueda}(2018)}]{Ashida18}%
  \BibitemOpen
  \bibfield  {author} {\bibinfo {author} {\bibfnamefont {Y.}~\bibnamefont
  {Ashida}}\ and\ \bibinfo {author} {\bibfnamefont {M.}~\bibnamefont {Ueda}},\
  }\href {\doibase 10.1103/PhysRevLett.120.185301} {\bibfield  {journal}
  {\bibinfo  {journal} {Phys. Rev. Lett.}\ }\textbf {\bibinfo {volume} {120}},\
  \bibinfo {pages} {185301} (\bibinfo {year} {2018})}\BibitemShut {NoStop}%
\bibitem [{\citenamefont {Yamamoto}\ \emph {et~al.}()\citenamefont {Yamamoto},
  \citenamefont {Nakagawa}, \citenamefont {Tsuji}, \citenamefont {Ueda},\ and\
  \citenamefont {Kawakami}}]{Yamamoto20}%
  \BibitemOpen
  \bibfield  {author} {\bibinfo {author} {\bibfnamefont {K.}~\bibnamefont
  {Yamamoto}}, \bibinfo {author} {\bibfnamefont {M.}~\bibnamefont {Nakagawa}},
  \bibinfo {author} {\bibfnamefont {N.}~\bibnamefont {Tsuji}}, \bibinfo
  {author} {\bibfnamefont {M.}~\bibnamefont {Ueda}}, \ and\ \bibinfo {author}
  {\bibfnamefont {N.}~\bibnamefont {Kawakami}},\ }\href@noop {} {\bibinfo
  {journal} {arXiv:2006.06169}\ }\BibitemShut {NoStop}%
\bibitem [{\citenamefont {Barreiro}\ \emph {et~al.}(2011)\citenamefont
  {Barreiro}, \citenamefont {M{\"u}ller}, \citenamefont {Schindler},
  \citenamefont {Nigg}, \citenamefont {Monz}, \citenamefont {Chwalla},
  \citenamefont {Hennrich}, \citenamefont {Roos}, \citenamefont {Zoller},\ and\
  \citenamefont {Blatt}}]{Barreiro11}%
  \BibitemOpen
\bibfield  {journal} {  }\bibfield  {author} {\bibinfo {author} {\bibfnamefont
  {J.~T.}\ \bibnamefont {Barreiro}}, \bibinfo {author} {\bibfnamefont
  {M.}~\bibnamefont {M{\"u}ller}}, \bibinfo {author} {\bibfnamefont
  {P.}~\bibnamefont {Schindler}}, \bibinfo {author} {\bibfnamefont
  {D.}~\bibnamefont {Nigg}}, \bibinfo {author} {\bibfnamefont {T.}~\bibnamefont
  {Monz}}, \bibinfo {author} {\bibfnamefont {M.}~\bibnamefont {Chwalla}},
  \bibinfo {author} {\bibfnamefont {M.}~\bibnamefont {Hennrich}}, \bibinfo
  {author} {\bibfnamefont {C.~F.}\ \bibnamefont {Roos}}, \bibinfo {author}
  {\bibfnamefont {P.}~\bibnamefont {Zoller}}, \ and\ \bibinfo {author}
  {\bibfnamefont {R.}~\bibnamefont {Blatt}},\ }\href@noop {} {\bibfield
  {journal} {\bibinfo  {journal} {Nature (London)}\ }\textbf {\bibinfo {volume}
  {470}},\ \bibinfo {pages} {486} (\bibinfo {year} {2011})}\BibitemShut
  {NoStop}%
\bibitem [{\citenamefont {Schindler}\ \emph {et~al.}(2013)\citenamefont
  {Schindler}, \citenamefont {M{\"u}ller}, \citenamefont {Nigg}, \citenamefont
  {Barreiro}, \citenamefont {Martinez}, \citenamefont {Hennrich}, \citenamefont
  {Monz}, \citenamefont {Diehl}, \citenamefont {Zoller},\ and\ \citenamefont
  {Blatt}}]{Schindler13}%
  \BibitemOpen
  \bibfield  {author} {\bibinfo {author} {\bibfnamefont {P.}~\bibnamefont
  {Schindler}}, \bibinfo {author} {\bibfnamefont {M.}~\bibnamefont
  {M{\"u}ller}}, \bibinfo {author} {\bibfnamefont {D.}~\bibnamefont {Nigg}},
  \bibinfo {author} {\bibfnamefont {J.~T.}\ \bibnamefont {Barreiro}}, \bibinfo
  {author} {\bibfnamefont {E.~A.}\ \bibnamefont {Martinez}}, \bibinfo {author}
  {\bibfnamefont {M.}~\bibnamefont {Hennrich}}, \bibinfo {author}
  {\bibfnamefont {T.}~\bibnamefont {Monz}}, \bibinfo {author} {\bibfnamefont
  {S.}~\bibnamefont {Diehl}}, \bibinfo {author} {\bibfnamefont
  {P.}~\bibnamefont {Zoller}}, \ and\ \bibinfo {author} {\bibfnamefont
  {R.}~\bibnamefont {Blatt}},\ }\href@noop {} {\bibfield  {journal} {\bibinfo
  {journal} {Nat. Phys.}\ }\textbf {\bibinfo {volume} {9}},\ \bibinfo {pages}
  {361} (\bibinfo {year} {2013})}\BibitemShut {NoStop}%
\bibitem [{\citenamefont {Liu}\ \emph {et~al.}(2011)\citenamefont {Liu},
  \citenamefont {Li}, \citenamefont {Huang}, \citenamefont {Li}, \citenamefont
  {Guo}, \citenamefont {Laine}, \citenamefont {Breuer},\ and\ \citenamefont
  {Piilo}}]{Liu11}%
  \BibitemOpen
  \bibfield  {author} {\bibinfo {author} {\bibfnamefont {B.-H.}\ \bibnamefont
  {Liu}}, \bibinfo {author} {\bibfnamefont {L.}~\bibnamefont {Li}}, \bibinfo
  {author} {\bibfnamefont {Y.-F.}\ \bibnamefont {Huang}}, \bibinfo {author}
  {\bibfnamefont {C.-F.}\ \bibnamefont {Li}}, \bibinfo {author} {\bibfnamefont
  {G.-C.}\ \bibnamefont {Guo}}, \bibinfo {author} {\bibfnamefont {E.-M.}\
  \bibnamefont {Laine}}, \bibinfo {author} {\bibfnamefont {H.-P.}\ \bibnamefont
  {Breuer}}, \ and\ \bibinfo {author} {\bibfnamefont {J.}~\bibnamefont
  {Piilo}},\ }\href@noop {} {\bibfield  {journal} {\bibinfo  {journal} {Nat.
  Phys.}\ }\textbf {\bibinfo {volume} {7}},\ \bibinfo {pages} {931} (\bibinfo
  {year} {2011})}\BibitemShut {NoStop}%
\bibitem [{\citenamefont {Liu}\ \emph {et~al.}(2018)\citenamefont {Liu},
  \citenamefont {Lyyra}, \citenamefont {Sun}, \citenamefont {Liu},
  \citenamefont {Li}, \citenamefont {Guo}, \citenamefont {Maniscalco},\ and\
  \citenamefont {Piilo}}]{Piilo18}%
  \BibitemOpen
  \bibfield  {author} {\bibinfo {author} {\bibfnamefont {Z.-D.}\ \bibnamefont
  {Liu}}, \bibinfo {author} {\bibfnamefont {H.}~\bibnamefont {Lyyra}}, \bibinfo
  {author} {\bibfnamefont {Y.-N.}\ \bibnamefont {Sun}}, \bibinfo {author}
  {\bibfnamefont {B.-H.}\ \bibnamefont {Liu}}, \bibinfo {author} {\bibfnamefont
  {C.-F.}\ \bibnamefont {Li}}, \bibinfo {author} {\bibfnamefont {G.-C.}\
  \bibnamefont {Guo}}, \bibinfo {author} {\bibfnamefont {S.}~\bibnamefont
  {Maniscalco}}, \ and\ \bibinfo {author} {\bibfnamefont {J.}~\bibnamefont
  {Piilo}},\ }\href@noop {} {\bibfield  {journal} {\bibinfo  {journal} {Nat.
  Commun.}\ }\textbf {\bibinfo {volume} {9}},\ \bibinfo {pages} {3453}
  (\bibinfo {year} {2018})}\BibitemShut {NoStop}%
\bibitem [{\citenamefont {Barontini}\ \emph {et~al.}(2013)\citenamefont
  {Barontini}, \citenamefont {Labouvie}, \citenamefont {Stubenrauch},
  \citenamefont {Vogler}, \citenamefont {Guarrera},\ and\ \citenamefont
  {Ott}}]{Ott13}%
  \BibitemOpen
  \bibfield  {author} {\bibinfo {author} {\bibfnamefont {G.}~\bibnamefont
  {Barontini}}, \bibinfo {author} {\bibfnamefont {R.}~\bibnamefont {Labouvie}},
  \bibinfo {author} {\bibfnamefont {F.}~\bibnamefont {Stubenrauch}}, \bibinfo
  {author} {\bibfnamefont {A.}~\bibnamefont {Vogler}}, \bibinfo {author}
  {\bibfnamefont {V.}~\bibnamefont {Guarrera}}, \ and\ \bibinfo {author}
  {\bibfnamefont {H.}~\bibnamefont {Ott}},\ }\href {\doibase
  10.1103/PhysRevLett.110.035302} {\bibfield  {journal} {\bibinfo  {journal}
  {Phys. Rev. Lett.}\ }\textbf {\bibinfo {volume} {110}},\ \bibinfo {pages}
  {035302} (\bibinfo {year} {2013})}\BibitemShut {NoStop}%
\bibitem [{\citenamefont {Labouvie}\ \emph {et~al.}(2016)\citenamefont
  {Labouvie}, \citenamefont {Santra}, \citenamefont {Heun},\ and\ \citenamefont
  {Ott}}]{Ott15}%
  \BibitemOpen
  \bibfield  {author} {\bibinfo {author} {\bibfnamefont {R.}~\bibnamefont
  {Labouvie}}, \bibinfo {author} {\bibfnamefont {B.}~\bibnamefont {Santra}},
  \bibinfo {author} {\bibfnamefont {S.}~\bibnamefont {Heun}}, \ and\ \bibinfo
  {author} {\bibfnamefont {H.}~\bibnamefont {Ott}},\ }\href {\doibase
  10.1103/PhysRevLett.116.235302} {\bibfield  {journal} {\bibinfo  {journal}
  {Phys. Rev. Lett.}\ }\textbf {\bibinfo {volume} {116}},\ \bibinfo {pages}
  {235302} (\bibinfo {year} {2016})}\BibitemShut {NoStop}%
\bibitem [{\citenamefont {Tomita}\ \emph {et~al.}(2017)\citenamefont {Tomita},
  \citenamefont {Nakajima}, \citenamefont {Danshita}, \citenamefont {Takasu},\
  and\ \citenamefont {Takahashi}}]{Tomita17}%
  \BibitemOpen
  \bibfield  {author} {\bibinfo {author} {\bibfnamefont {T.}~\bibnamefont
  {Tomita}}, \bibinfo {author} {\bibfnamefont {S.}~\bibnamefont {Nakajima}},
  \bibinfo {author} {\bibfnamefont {I.}~\bibnamefont {Danshita}}, \bibinfo
  {author} {\bibfnamefont {Y.}~\bibnamefont {Takasu}}, \ and\ \bibinfo {author}
  {\bibfnamefont {Y.}~\bibnamefont {Takahashi}},\ }\href@noop {} {\bibfield
  {journal} {\bibinfo  {journal} {Sci. Adv.}\ }\textbf {\bibinfo {volume}
  {3}},\ \bibinfo {pages} {e1701513} (\bibinfo {year} {2017})}\BibitemShut
  {NoStop}%
\bibitem [{\citenamefont {Sponselee}\ \emph {et~al.}(2018)\citenamefont
  {Sponselee}, \citenamefont {Freystatzky}, \citenamefont {Abeln},
  \citenamefont {Diem}, \citenamefont {Hundt}, \citenamefont {Kochanke},
  \citenamefont {Ponath}, \citenamefont {Santra}, \citenamefont {Mathey},
  \citenamefont {Sengstock},\ and\ \citenamefont {Becker}}]{Spon18}%
  \BibitemOpen
  \bibfield  {author} {\bibinfo {author} {\bibfnamefont {K.}~\bibnamefont
  {Sponselee}}, \bibinfo {author} {\bibfnamefont {L.}~\bibnamefont
  {Freystatzky}}, \bibinfo {author} {\bibfnamefont {B.}~\bibnamefont {Abeln}},
  \bibinfo {author} {\bibfnamefont {M.}~\bibnamefont {Diem}}, \bibinfo {author}
  {\bibfnamefont {B.}~\bibnamefont {Hundt}}, \bibinfo {author} {\bibfnamefont
  {A.}~\bibnamefont {Kochanke}}, \bibinfo {author} {\bibfnamefont
  {T.}~\bibnamefont {Ponath}}, \bibinfo {author} {\bibfnamefont
  {B.}~\bibnamefont {Santra}}, \bibinfo {author} {\bibfnamefont
  {L.}~\bibnamefont {Mathey}}, \bibinfo {author} {\bibfnamefont
  {K.}~\bibnamefont {Sengstock}}, \ and\ \bibinfo {author} {\bibfnamefont
  {C.}~\bibnamefont {Becker}},\ }\href@noop {} {\bibfield  {journal} {\bibinfo
  {journal} {Quantum Sci. Technol.}\ }\textbf {\bibinfo {volume} {4}},\
  \bibinfo {pages} {014002} (\bibinfo {year} {2018})}\BibitemShut {NoStop}%
\bibitem [{\citenamefont {Tomita}\ \emph {et~al.}(2019)\citenamefont {Tomita},
  \citenamefont {Nakajima}, \citenamefont {Takasu},\ and\ \citenamefont
  {Takahashi}}]{Tomita19}%
  \BibitemOpen
  \bibfield  {author} {\bibinfo {author} {\bibfnamefont {T.}~\bibnamefont
  {Tomita}}, \bibinfo {author} {\bibfnamefont {S.}~\bibnamefont {Nakajima}},
  \bibinfo {author} {\bibfnamefont {Y.}~\bibnamefont {Takasu}}, \ and\ \bibinfo
  {author} {\bibfnamefont {Y.}~\bibnamefont {Takahashi}},\ }\href {\doibase
  10.1103/PhysRevA.99.031601} {\bibfield  {journal} {\bibinfo  {journal} {Phys.
  Rev. A}\ }\textbf {\bibinfo {volume} {99}},\ \bibinfo {pages} {031601(R)}
  (\bibinfo {year} {2019})}\BibitemShut {NoStop}%
\bibitem [{\citenamefont {Bouganne}\ \emph {et~al.}(2019)\citenamefont
  {Bouganne}, \citenamefont {Aguilera}, \citenamefont {Ghermaoui},
  \citenamefont {Beugnon},\ and\ \citenamefont {Gerbier}}]{Gerbier19}%
  \BibitemOpen
  \bibfield  {author} {\bibinfo {author} {\bibfnamefont {R.}~\bibnamefont
  {Bouganne}}, \bibinfo {author} {\bibfnamefont {M.~B.}\ \bibnamefont
  {Aguilera}}, \bibinfo {author} {\bibfnamefont {A.}~\bibnamefont {Ghermaoui}},
  \bibinfo {author} {\bibfnamefont {J.}~\bibnamefont {Beugnon}}, \ and\
  \bibinfo {author} {\bibfnamefont {F.}~\bibnamefont {Gerbier}},\ }\href@noop
  {} {\bibfield  {journal} {\bibinfo  {journal} {Nat. Phys.}\ }\textbf
  {\bibinfo {volume} {16}},\ \bibinfo {pages} {21} (\bibinfo {year}
  {2019})}\BibitemShut {NoStop}%
\bibitem [{\citenamefont {Takasu}\ \emph {et~al.}()\citenamefont {Takasu},
  \citenamefont {Yagami}, \citenamefont {Ashida}, \citenamefont {Hamazaki},
  \citenamefont {Kuno},\ and\ \citenamefont {Takahashi}}]{Takasu20}%
  \BibitemOpen
  \bibfield  {author} {\bibinfo {author} {\bibfnamefont {Y.}~\bibnamefont
  {Takasu}}, \bibinfo {author} {\bibfnamefont {T.}~\bibnamefont {Yagami}},
  \bibinfo {author} {\bibfnamefont {Y.}~\bibnamefont {Ashida}}, \bibinfo
  {author} {\bibfnamefont {R.}~\bibnamefont {Hamazaki}}, \bibinfo {author}
  {\bibfnamefont {Y.}~\bibnamefont {Kuno}}, \ and\ \bibinfo {author}
  {\bibfnamefont {Y.}~\bibnamefont {Takahashi}},\ }\href@noop {} {\bibinfo
  {journal} {arXiv:2004.05734}\ }\BibitemShut {NoStop}%
\bibitem [{\citenamefont {Kasprzak}\ \emph {et~al.}(2006)\citenamefont
  {Kasprzak}, \citenamefont {Richard}, \citenamefont {Kundermann},
  \citenamefont {Baas}, \citenamefont {Jeambrun}, \citenamefont {Keeling},
  \citenamefont {Marchetti}, \citenamefont {Szyma{\'n}ska}, \citenamefont
  {Andr{\'e}}, \citenamefont {Staehli}, \citenamefont {Savona}, \citenamefont
  {Littlewood}, \citenamefont {Deveaud},\ and\ \citenamefont
  {Dang}}]{Kasprzak06}%
  \BibitemOpen
\bibfield  {journal} {  }\bibfield  {author} {\bibinfo {author} {\bibfnamefont
  {J.}~\bibnamefont {Kasprzak}}, \bibinfo {author} {\bibfnamefont
  {M.}~\bibnamefont {Richard}}, \bibinfo {author} {\bibfnamefont
  {S.}~\bibnamefont {Kundermann}}, \bibinfo {author} {\bibfnamefont
  {A.}~\bibnamefont {Baas}}, \bibinfo {author} {\bibfnamefont {P.}~\bibnamefont
  {Jeambrun}}, \bibinfo {author} {\bibfnamefont {J.~M.~J.}\ \bibnamefont
  {Keeling}}, \bibinfo {author} {\bibfnamefont {F.~M.}\ \bibnamefont
  {Marchetti}}, \bibinfo {author} {\bibfnamefont {M.~H.}\ \bibnamefont
  {Szyma{\'n}ska}}, \bibinfo {author} {\bibfnamefont {R.}~\bibnamefont
  {Andr{\'e}}}, \bibinfo {author} {\bibfnamefont {J.~L.}\ \bibnamefont
  {Staehli}}, \bibinfo {author} {\bibfnamefont {V.}~\bibnamefont {Savona}},
  \bibinfo {author} {\bibfnamefont {P.~B.}\ \bibnamefont {Littlewood}},
  \bibinfo {author} {\bibfnamefont {B.}~\bibnamefont {Deveaud}}, \ and\
  \bibinfo {author} {\bibfnamefont {L.~S.}\ \bibnamefont {Dang}},\ }\href@noop
  {} {\bibfield  {journal} {\bibinfo  {journal} {Nature (London)}\ }\textbf
  {\bibinfo {volume} {443}},\ \bibinfo {pages} {409} (\bibinfo {year}
  {2006})}\BibitemShut {NoStop}%
\bibitem [{\citenamefont {Leyder}\ \emph {et~al.}(2007)\citenamefont {Leyder},
  \citenamefont {Romanelli}, \citenamefont {Karr}, \citenamefont {Giacobino},
  \citenamefont {Liew}, \citenamefont {Glazov}, \citenamefont {Kavokin},
  \citenamefont {Malpuech},\ and\ \citenamefont {Bramati}}]{Leyder07}%
  \BibitemOpen
  \bibfield  {author} {\bibinfo {author} {\bibfnamefont {C.}~\bibnamefont
  {Leyder}}, \bibinfo {author} {\bibfnamefont {M.}~\bibnamefont {Romanelli}},
  \bibinfo {author} {\bibfnamefont {J.~P.}\ \bibnamefont {Karr}}, \bibinfo
  {author} {\bibfnamefont {E.}~\bibnamefont {Giacobino}}, \bibinfo {author}
  {\bibfnamefont {T.~C.}\ \bibnamefont {Liew}}, \bibinfo {author}
  {\bibfnamefont {M.~M.}\ \bibnamefont {Glazov}}, \bibinfo {author}
  {\bibfnamefont {A.~V.}\ \bibnamefont {Kavokin}}, \bibinfo {author}
  {\bibfnamefont {G.}~\bibnamefont {Malpuech}}, \ and\ \bibinfo {author}
  {\bibfnamefont {A.}~\bibnamefont {Bramati}},\ }\href@noop {} {\bibfield
  {journal} {\bibinfo  {journal} {Nat. Phys.}\ }\textbf {\bibinfo {volume}
  {3}},\ \bibinfo {pages} {628} (\bibinfo {year} {2007})}\BibitemShut {NoStop}%
\bibitem [{\citenamefont {Byrnes}\ \emph {et~al.}(2014)\citenamefont {Byrnes},
  \citenamefont {Kim},\ and\ \citenamefont {Yamamoto}}]{Byrnes14}%
  \BibitemOpen
  \bibfield  {author} {\bibinfo {author} {\bibfnamefont {T.}~\bibnamefont
  {Byrnes}}, \bibinfo {author} {\bibfnamefont {N.~Y.}\ \bibnamefont {Kim}}, \
  and\ \bibinfo {author} {\bibfnamefont {Y.}~\bibnamefont {Yamamoto}},\
  }\href@noop {} {\bibfield  {journal} {\bibinfo  {journal} {Nat. Phys.}\
  }\textbf {\bibinfo {volume} {10}},\ \bibinfo {pages} {803} (\bibinfo {year}
  {2014})}\BibitemShut {NoStop}%
\bibitem [{\citenamefont {Gao}\ \emph {et~al.}(2015)\citenamefont {Gao},
  \citenamefont {Estrecho}, \citenamefont {Bliokh}, \citenamefont {Liew},
  \citenamefont {Fraser}, \citenamefont {Brodbeck}, \citenamefont {Kamp},
  \citenamefont {Schneider}, \citenamefont {H{\"o}fling}, \citenamefont
  {Yamamoto}, \citenamefont {Nori}, \citenamefont {Truscott}, \citenamefont
  {Dall},\ and\ \citenamefont {Ostrovskaya}}]{Gao15}%
  \BibitemOpen
  \bibfield  {author} {\bibinfo {author} {\bibfnamefont {T.}~\bibnamefont
  {Gao}}, \bibinfo {author} {\bibfnamefont {E.}~\bibnamefont {Estrecho}},
  \bibinfo {author} {\bibfnamefont {K.~Y.}\ \bibnamefont {Bliokh}}, \bibinfo
  {author} {\bibfnamefont {T.~C.~H.}\ \bibnamefont {Liew}}, \bibinfo {author}
  {\bibfnamefont {M.~D.}\ \bibnamefont {Fraser}}, \bibinfo {author}
  {\bibfnamefont {S.}~\bibnamefont {Brodbeck}}, \bibinfo {author}
  {\bibfnamefont {M.}~\bibnamefont {Kamp}}, \bibinfo {author} {\bibfnamefont
  {C.}~\bibnamefont {Schneider}}, \bibinfo {author} {\bibfnamefont
  {S.}~\bibnamefont {H{\"o}fling}}, \bibinfo {author} {\bibfnamefont
  {Y.}~\bibnamefont {Yamamoto}}, \bibinfo {author} {\bibfnamefont
  {F.}~\bibnamefont {Nori}}, \bibinfo {author} {\bibfnamefont {A.~G.}\
  \bibnamefont {Truscott}}, \bibinfo {author} {\bibfnamefont {R.~G.}\
  \bibnamefont {Dall}}, \ and\ \bibinfo {author} {\bibfnamefont {E.~A.}\
  \bibnamefont {Ostrovskaya}},\ }\href@noop {} {\bibfield  {journal} {\bibinfo
  {journal} {Nature (London)}\ }\textbf {\bibinfo {volume} {526}},\ \bibinfo
  {pages} {554} (\bibinfo {year} {2015})}\BibitemShut {NoStop}%
\bibitem [{\citenamefont {Baboux}\ \emph {et~al.}(2016)\citenamefont {Baboux},
  \citenamefont {Ge}, \citenamefont {Jacqmin}, \citenamefont {Biondi},
  \citenamefont {Galopin}, \citenamefont {Lema\^{\i}tre}, \citenamefont
  {Le~Gratiet}, \citenamefont {Sagnes}, \citenamefont {Schmidt}, \citenamefont
  {T\"ureci}, \citenamefont {Amo},\ and\ \citenamefont {Bloch}}]{Baboux16}%
  \BibitemOpen
  \bibfield  {author} {\bibinfo {author} {\bibfnamefont {F.}~\bibnamefont
  {Baboux}}, \bibinfo {author} {\bibfnamefont {L.}~\bibnamefont {Ge}}, \bibinfo
  {author} {\bibfnamefont {T.}~\bibnamefont {Jacqmin}}, \bibinfo {author}
  {\bibfnamefont {M.}~\bibnamefont {Biondi}}, \bibinfo {author} {\bibfnamefont
  {E.}~\bibnamefont {Galopin}}, \bibinfo {author} {\bibfnamefont
  {A.}~\bibnamefont {Lema\^{\i}tre}}, \bibinfo {author} {\bibfnamefont
  {L.}~\bibnamefont {Le~Gratiet}}, \bibinfo {author} {\bibfnamefont
  {I.}~\bibnamefont {Sagnes}}, \bibinfo {author} {\bibfnamefont
  {S.}~\bibnamefont {Schmidt}}, \bibinfo {author} {\bibfnamefont {H.~E.}\
  \bibnamefont {T\"ureci}}, \bibinfo {author} {\bibfnamefont {A.}~\bibnamefont
  {Amo}}, \ and\ \bibinfo {author} {\bibfnamefont {J.}~\bibnamefont {Bloch}},\
  }\href {\doibase 10.1103/PhysRevLett.116.066402} {\bibfield  {journal}
  {\bibinfo  {journal} {Phys. Rev. Lett.}\ }\textbf {\bibinfo {volume} {116}},\
  \bibinfo {pages} {066402} (\bibinfo {year} {2016})}\BibitemShut {NoStop}%
\bibitem [{\citenamefont {Klembt}\ \emph {et~al.}(2018)\citenamefont {Klembt},
  \citenamefont {Harder}, \citenamefont {Egorov}, \citenamefont {Winkler},
  \citenamefont {Ge}, \citenamefont {Bandres}, \citenamefont {Emmerling},
  \citenamefont {Worschech}, \citenamefont {Liew}, \citenamefont {Segev},
  \citenamefont {Schneider},\ and\ \citenamefont {H{\"o}fling}}]{Klembt18}%
  \BibitemOpen
  \bibfield  {author} {\bibinfo {author} {\bibfnamefont {S.}~\bibnamefont
  {Klembt}}, \bibinfo {author} {\bibfnamefont {T.}~\bibnamefont {Harder}},
  \bibinfo {author} {\bibfnamefont {O.}~\bibnamefont {Egorov}}, \bibinfo
  {author} {\bibfnamefont {K.}~\bibnamefont {Winkler}}, \bibinfo {author}
  {\bibfnamefont {R.}~\bibnamefont {Ge}}, \bibinfo {author} {\bibfnamefont
  {M.}~\bibnamefont {Bandres}}, \bibinfo {author} {\bibfnamefont
  {M.}~\bibnamefont {Emmerling}}, \bibinfo {author} {\bibfnamefont
  {L.}~\bibnamefont {Worschech}}, \bibinfo {author} {\bibfnamefont
  {T.}~\bibnamefont {Liew}}, \bibinfo {author} {\bibfnamefont {M.}~\bibnamefont
  {Segev}}, \bibinfo {author} {\bibfnamefont {C.}~\bibnamefont {Schneider}}, \
  and\ \bibinfo {author} {\bibfnamefont {S.}~\bibnamefont {H{\"o}fling}},\
  }\href@noop {} {\bibfield  {journal} {\bibinfo  {journal} {Nature (London)}\
  }\textbf {\bibinfo {volume} {562}},\ \bibinfo {pages} {552} (\bibinfo {year}
  {2018})}\BibitemShut {NoStop}%
\bibitem [{\citenamefont {Frensley}(1990)}]{William90}%
  \BibitemOpen
  \bibfield  {author} {\bibinfo {author} {\bibfnamefont {W.~R.}\ \bibnamefont
  {Frensley}},\ }\href {\doibase 10.1103/RevModPhys.62.745} {\bibfield
  {journal} {\bibinfo  {journal} {Rev. Mod. Phys.}\ }\textbf {\bibinfo {volume}
  {62}},\ \bibinfo {pages} {745} (\bibinfo {year} {1990})}\BibitemShut
  {NoStop}%
\bibitem [{\citenamefont {Tokura}\ and\ \citenamefont
  {Nagaosa}(2018)}]{Tokura18}%
  \BibitemOpen
  \bibfield  {author} {\bibinfo {author} {\bibfnamefont {Y.}~\bibnamefont
  {Tokura}}\ and\ \bibinfo {author} {\bibfnamefont {N.}~\bibnamefont
  {Nagaosa}},\ }\href@noop {} {\bibfield  {journal} {\bibinfo  {journal} {Nat.
  Commun.}\ }\textbf {\bibinfo {volume} {9}},\ \bibinfo {pages} {3740}
  (\bibinfo {year} {2018})}\BibitemShut {NoStop}%
\bibitem [{\citenamefont {Rikken}\ and\ \citenamefont
  {Raupach}(1997)}]{Rikken97}%
  \BibitemOpen
  \bibfield  {author} {\bibinfo {author} {\bibfnamefont {G.}~\bibnamefont
  {Rikken}}\ and\ \bibinfo {author} {\bibfnamefont {E.}~\bibnamefont
  {Raupach}},\ }\href@noop {} {\bibfield  {journal} {\bibinfo  {journal}
  {Nature (London)}\ }\textbf {\bibinfo {volume} {390}},\ \bibinfo {pages}
  {493} (\bibinfo {year} {1997})}\BibitemShut {NoStop}%
\bibitem [{\citenamefont {Linke}\ \emph {et~al.}(1999)\citenamefont {Linke},
  \citenamefont {Humphrey}, \citenamefont {L{\"o}fgren}, \citenamefont
  {Sushkov}, \citenamefont {Newbury}, \citenamefont {Taylor},\ and\
  \citenamefont {Omling}}]{Linke99}%
  \BibitemOpen
  \bibfield  {author} {\bibinfo {author} {\bibfnamefont {H.}~\bibnamefont
  {Linke}}, \bibinfo {author} {\bibfnamefont {T.}~\bibnamefont {Humphrey}},
  \bibinfo {author} {\bibfnamefont {A.}~\bibnamefont {L{\"o}fgren}}, \bibinfo
  {author} {\bibfnamefont {A.}~\bibnamefont {Sushkov}}, \bibinfo {author}
  {\bibfnamefont {R.}~\bibnamefont {Newbury}}, \bibinfo {author} {\bibfnamefont
  {R.}~\bibnamefont {Taylor}}, \ and\ \bibinfo {author} {\bibfnamefont
  {P.}~\bibnamefont {Omling}},\ }\href@noop {} {\bibfield  {journal} {\bibinfo
  {journal} {Science}\ }\textbf {\bibinfo {volume} {286}},\ \bibinfo {pages}
  {2314} (\bibinfo {year} {1999})}\BibitemShut {NoStop}%
\bibitem [{\citenamefont {Morimoto}\ and\ \citenamefont
  {Nagaosa}(2016{\natexlab{a}})}]{Morimoto16}%
  \BibitemOpen
  \bibfield  {author} {\bibinfo {author} {\bibfnamefont {T.}~\bibnamefont
  {Morimoto}}\ and\ \bibinfo {author} {\bibfnamefont {N.}~\bibnamefont
  {Nagaosa}},\ }\href@noop {} {\bibfield  {journal} {\bibinfo  {journal} {Sci.
  Adv.}\ }\textbf {\bibinfo {volume} {2}},\ \bibinfo {pages} {e1501524}
  (\bibinfo {year} {2016}{\natexlab{a}})}\BibitemShut {NoStop}%
\bibitem [{\citenamefont {Kitamura}\ \emph {et~al.}(2020)\citenamefont
  {Kitamura}, \citenamefont {Nagaosa},\ and\ \citenamefont
  {Morimoto}}]{Kitamura20}%
  \BibitemOpen
  \bibfield  {author} {\bibinfo {author} {\bibfnamefont {S.}~\bibnamefont
  {Kitamura}}, \bibinfo {author} {\bibfnamefont {N.}~\bibnamefont {Nagaosa}}, \
  and\ \bibinfo {author} {\bibfnamefont {T.}~\bibnamefont {Morimoto}},\
  }\href@noop {} {\bibfield  {journal} {\bibinfo  {journal} {Commun. Phys.}\
  }\textbf {\bibinfo {volume} {3}},\ \bibinfo {pages} {63} (\bibinfo {year}
  {2020})}\BibitemShut {NoStop}%
\bibitem [{\citenamefont {Ono}\ \emph {et~al.}(2002)\citenamefont {Ono},
  \citenamefont {Austing}, \citenamefont {Tokura},\ and\ \citenamefont
  {Tarucha}}]{Ono02}%
  \BibitemOpen
  \bibfield  {author} {\bibinfo {author} {\bibfnamefont {K.}~\bibnamefont
  {Ono}}, \bibinfo {author} {\bibfnamefont {D.}~\bibnamefont {Austing}},
  \bibinfo {author} {\bibfnamefont {Y.}~\bibnamefont {Tokura}}, \ and\ \bibinfo
  {author} {\bibfnamefont {S.}~\bibnamefont {Tarucha}},\ }\href@noop {}
  {\bibfield  {journal} {\bibinfo  {journal} {Science}\ }\textbf {\bibinfo
  {volume} {297}},\ \bibinfo {pages} {1313} (\bibinfo {year}
  {2002})}\BibitemShut {NoStop}%
\bibitem [{\citenamefont {Scheibner}\ \emph {et~al.}(2008)\citenamefont
  {Scheibner}, \citenamefont {K{\"o}nig}, \citenamefont {Reuter}, \citenamefont
  {Wieck}, \citenamefont {Gould}, \citenamefont {Buhmann},\ and\ \citenamefont
  {Molenkamp}}]{Molenkamp08}%
  \BibitemOpen
  \bibfield  {author} {\bibinfo {author} {\bibfnamefont {R.}~\bibnamefont
  {Scheibner}}, \bibinfo {author} {\bibfnamefont {M.}~\bibnamefont
  {K{\"o}nig}}, \bibinfo {author} {\bibfnamefont {D.}~\bibnamefont {Reuter}},
  \bibinfo {author} {\bibfnamefont {A.}~\bibnamefont {Wieck}}, \bibinfo
  {author} {\bibfnamefont {C.}~\bibnamefont {Gould}}, \bibinfo {author}
  {\bibfnamefont {H.}~\bibnamefont {Buhmann}}, \ and\ \bibinfo {author}
  {\bibfnamefont {L.}~\bibnamefont {Molenkamp}},\ }\href@noop {} {\bibfield
  {journal} {\bibinfo  {journal} {New J. Phys.}\ }\textbf {\bibinfo {volume}
  {10}},\ \bibinfo {pages} {083016} (\bibinfo {year} {2008})}\BibitemShut
  {NoStop}%
\bibitem [{\citenamefont {Ramezani}\ \emph {et~al.}(2010)\citenamefont
  {Ramezani}, \citenamefont {Kottos}, \citenamefont {El-Ganainy},\ and\
  \citenamefont {Christodoulides}}]{Hamidreza10}%
  \BibitemOpen
  \bibfield  {author} {\bibinfo {author} {\bibfnamefont {H.}~\bibnamefont
  {Ramezani}}, \bibinfo {author} {\bibfnamefont {T.}~\bibnamefont {Kottos}},
  \bibinfo {author} {\bibfnamefont {R.}~\bibnamefont {El-Ganainy}}, \ and\
  \bibinfo {author} {\bibfnamefont {D.~N.}\ \bibnamefont {Christodoulides}},\
  }\href {\doibase 10.1103/PhysRevA.82.043803} {\bibfield  {journal} {\bibinfo
  {journal} {Phys. Rev. A}\ }\textbf {\bibinfo {volume} {82}},\ \bibinfo
  {pages} {043803} (\bibinfo {year} {2010})}\BibitemShut {NoStop}%
\bibitem [{\citenamefont {Chang}\ \emph {et~al.}(2014)\citenamefont {Chang},
  \citenamefont {Jiang}, \citenamefont {Hua}, \citenamefont {Yang},
  \citenamefont {Wen}, \citenamefont {Jiang}, \citenamefont {Li}, \citenamefont
  {Wang},\ and\ \citenamefont {Xiao}}]{Chang14}%
  \BibitemOpen
  \bibfield  {author} {\bibinfo {author} {\bibfnamefont {L.}~\bibnamefont
  {Chang}}, \bibinfo {author} {\bibfnamefont {X.}~\bibnamefont {Jiang}},
  \bibinfo {author} {\bibfnamefont {S.}~\bibnamefont {Hua}}, \bibinfo {author}
  {\bibfnamefont {C.}~\bibnamefont {Yang}}, \bibinfo {author} {\bibfnamefont
  {J.}~\bibnamefont {Wen}}, \bibinfo {author} {\bibfnamefont {L.}~\bibnamefont
  {Jiang}}, \bibinfo {author} {\bibfnamefont {G.}~\bibnamefont {Li}}, \bibinfo
  {author} {\bibfnamefont {G.}~\bibnamefont {Wang}}, \ and\ \bibinfo {author}
  {\bibfnamefont {M.}~\bibnamefont {Xiao}},\ }\href@noop {} {\bibfield
  {journal} {\bibinfo  {journal} {Nat. Photon.}\ }\textbf {\bibinfo {volume}
  {8}},\ \bibinfo {pages} {524} (\bibinfo {year} {2014})}\BibitemShut {NoStop}%
\bibitem [{\citenamefont {Peng}\ \emph {et~al.}(2014)\citenamefont {Peng},
  \citenamefont {{\"O}zdemir}, \citenamefont {Lei}, \citenamefont {Monifi},
  \citenamefont {Gianfreda}, \citenamefont {Long}, \citenamefont {Fan},
  \citenamefont {Nori}, \citenamefont {Bender},\ and\ \citenamefont
  {Yang}}]{Peng14}%
  \BibitemOpen
  \bibfield  {author} {\bibinfo {author} {\bibfnamefont {B.}~\bibnamefont
  {Peng}}, \bibinfo {author} {\bibfnamefont {{\c{S}}.~K.}\ \bibnamefont
  {{\"O}zdemir}}, \bibinfo {author} {\bibfnamefont {F.}~\bibnamefont {Lei}},
  \bibinfo {author} {\bibfnamefont {F.}~\bibnamefont {Monifi}}, \bibinfo
  {author} {\bibfnamefont {M.}~\bibnamefont {Gianfreda}}, \bibinfo {author}
  {\bibfnamefont {G.~L.}\ \bibnamefont {Long}}, \bibinfo {author}
  {\bibfnamefont {S.}~\bibnamefont {Fan}}, \bibinfo {author} {\bibfnamefont
  {F.}~\bibnamefont {Nori}}, \bibinfo {author} {\bibfnamefont {C.~M.}\
  \bibnamefont {Bender}}, \ and\ \bibinfo {author} {\bibfnamefont
  {L.}~\bibnamefont {Yang}},\ }\href@noop {} {\bibfield  {journal} {\bibinfo
  {journal} {Nat. Phys.}\ }\textbf {\bibinfo {volume} {10}},\ \bibinfo {pages}
  {394} (\bibinfo {year} {2014})}\BibitemShut {NoStop}%
\bibitem [{\citenamefont {Nazari}\ \emph {et~al.}(2014)\citenamefont {Nazari},
  \citenamefont {Bender}, \citenamefont {Ramezani}, \citenamefont
  {Moravvej-Farshi}, \citenamefont {Christodoulides},\ and\ \citenamefont
  {Kottos}}]{Nazari14}%
  \BibitemOpen
  \bibfield  {author} {\bibinfo {author} {\bibfnamefont {F.}~\bibnamefont
  {Nazari}}, \bibinfo {author} {\bibfnamefont {N.}~\bibnamefont {Bender}},
  \bibinfo {author} {\bibfnamefont {H.}~\bibnamefont {Ramezani}}, \bibinfo
  {author} {\bibfnamefont {M.}~\bibnamefont {Moravvej-Farshi}}, \bibinfo
  {author} {\bibfnamefont {D.}~\bibnamefont {Christodoulides}}, \ and\ \bibinfo
  {author} {\bibfnamefont {T.}~\bibnamefont {Kottos}},\ }\href@noop {}
  {\bibfield  {journal} {\bibinfo  {journal} {Opt. Express}\ }\textbf {\bibinfo
  {volume} {22}},\ \bibinfo {pages} {9574} (\bibinfo {year}
  {2014})}\BibitemShut {NoStop}%
\bibitem [{\citenamefont {Lanneb{\`e}re}\ and\ \citenamefont
  {Silveirinha}(2015)}]{Sylvain15}%
  \BibitemOpen
  \bibfield  {author} {\bibinfo {author} {\bibfnamefont {S.}~\bibnamefont
  {Lanneb{\`e}re}}\ and\ \bibinfo {author} {\bibfnamefont {M.~G.}\ \bibnamefont
  {Silveirinha}},\ }\href@noop {} {\bibfield  {journal} {\bibinfo  {journal}
  {Nat. Commun.}\ }\textbf {\bibinfo {volume} {6}},\ \bibinfo {pages} {8766}
  (\bibinfo {year} {2015})}\BibitemShut {NoStop}%
\bibitem [{\citenamefont {Liang}\ \emph {et~al.}(2009)\citenamefont {Liang},
  \citenamefont {Yuan},\ and\ \citenamefont {Cheng}}]{Liang09}%
  \BibitemOpen
  \bibfield  {author} {\bibinfo {author} {\bibfnamefont {B.}~\bibnamefont
  {Liang}}, \bibinfo {author} {\bibfnamefont {B.}~\bibnamefont {Yuan}}, \ and\
  \bibinfo {author} {\bibfnamefont {J.-c.}\ \bibnamefont {Cheng}},\ }\href
  {\doibase 10.1103/PhysRevLett.103.104301} {\bibfield  {journal} {\bibinfo
  {journal} {Phys. Rev. Lett.}\ }\textbf {\bibinfo {volume} {103}},\ \bibinfo
  {pages} {104301} (\bibinfo {year} {2009})}\BibitemShut {NoStop}%
\bibitem [{\citenamefont {Fleury}\ \emph {et~al.}(2014)\citenamefont {Fleury},
  \citenamefont {Sounas}, \citenamefont {Sieck}, \citenamefont {Haberman},\
  and\ \citenamefont {Al{\`u}}}]{Andrea14}%
  \BibitemOpen
  \bibfield  {author} {\bibinfo {author} {\bibfnamefont {R.}~\bibnamefont
  {Fleury}}, \bibinfo {author} {\bibfnamefont {D.~L.}\ \bibnamefont {Sounas}},
  \bibinfo {author} {\bibfnamefont {C.~F.}\ \bibnamefont {Sieck}}, \bibinfo
  {author} {\bibfnamefont {M.~R.}\ \bibnamefont {Haberman}}, \ and\ \bibinfo
  {author} {\bibfnamefont {A.}~\bibnamefont {Al{\`u}}},\ }\href@noop {}
  {\bibfield  {journal} {\bibinfo  {journal} {Science}\ }\textbf {\bibinfo
  {volume} {343}},\ \bibinfo {pages} {516} (\bibinfo {year}
  {2014})}\BibitemShut {NoStop}%
\bibitem [{\citenamefont {Li}\ \emph {et~al.}(2014)\citenamefont {Li},
  \citenamefont {Anzel}, \citenamefont {Yang}, \citenamefont {Kevrekidis},\
  and\ \citenamefont {Daraio}}]{Li14}%
  \BibitemOpen
  \bibfield  {author} {\bibinfo {author} {\bibfnamefont {F.}~\bibnamefont
  {Li}}, \bibinfo {author} {\bibfnamefont {P.}~\bibnamefont {Anzel}}, \bibinfo
  {author} {\bibfnamefont {J.}~\bibnamefont {Yang}}, \bibinfo {author}
  {\bibfnamefont {P.~G.}\ \bibnamefont {Kevrekidis}}, \ and\ \bibinfo {author}
  {\bibfnamefont {C.}~\bibnamefont {Daraio}},\ }\href@noop {} {\bibfield
  {journal} {\bibinfo  {journal} {Nat. Commun.}\ }\textbf {\bibinfo {volume}
  {5}},\ \bibinfo {pages} {5311} (\bibinfo {year} {2014})}\BibitemShut
  {NoStop}%
\bibitem [{\citenamefont {Wang}\ \emph {et~al.}(2015)\citenamefont {Wang},
  \citenamefont {Yang}, \citenamefont {Ni}, \citenamefont {Xu}, \citenamefont
  {Sun}, \citenamefont {Chen}, \citenamefont {Feng}, \citenamefont {Liu},
  \citenamefont {Lu},\ and\ \citenamefont {Chen}}]{Chen15}%
  \BibitemOpen
  \bibfield  {author} {\bibinfo {author} {\bibfnamefont {Q.}~\bibnamefont
  {Wang}}, \bibinfo {author} {\bibfnamefont {Y.}~\bibnamefont {Yang}}, \bibinfo
  {author} {\bibfnamefont {X.}~\bibnamefont {Ni}}, \bibinfo {author}
  {\bibfnamefont {Y.-L.}\ \bibnamefont {Xu}}, \bibinfo {author} {\bibfnamefont
  {X.-C.}\ \bibnamefont {Sun}}, \bibinfo {author} {\bibfnamefont {Z.-G.}\
  \bibnamefont {Chen}}, \bibinfo {author} {\bibfnamefont {L.}~\bibnamefont
  {Feng}}, \bibinfo {author} {\bibfnamefont {X.-p.}\ \bibnamefont {Liu}},
  \bibinfo {author} {\bibfnamefont {M.-H.}\ \bibnamefont {Lu}}, \ and\ \bibinfo
  {author} {\bibfnamefont {Y.-F.}\ \bibnamefont {Chen}},\ }\href@noop {}
  {\bibfield  {journal} {\bibinfo  {journal} {Sci. Rep.}\ }\textbf {\bibinfo
  {volume} {5}},\ \bibinfo {pages} {1} (\bibinfo {year} {2015})}\BibitemShut
  {NoStop}%
\bibitem [{\citenamefont {Wang}\ \emph {et~al.}(2018)\citenamefont {Wang},
  \citenamefont {Yousefzadeh}, \citenamefont {Chen}, \citenamefont {Nassar},
  \citenamefont {Huang},\ and\ \citenamefont {Daraio}}]{Daraio18}%
  \BibitemOpen
  \bibfield  {author} {\bibinfo {author} {\bibfnamefont {Y.}~\bibnamefont
  {Wang}}, \bibinfo {author} {\bibfnamefont {B.}~\bibnamefont {Yousefzadeh}},
  \bibinfo {author} {\bibfnamefont {H.}~\bibnamefont {Chen}}, \bibinfo {author}
  {\bibfnamefont {H.}~\bibnamefont {Nassar}}, \bibinfo {author} {\bibfnamefont
  {G.}~\bibnamefont {Huang}}, \ and\ \bibinfo {author} {\bibfnamefont
  {C.}~\bibnamefont {Daraio}},\ }\href {\doibase
  10.1103/PhysRevLett.121.194301} {\bibfield  {journal} {\bibinfo  {journal}
  {Phys. Rev. Lett.}\ }\textbf {\bibinfo {volume} {121}},\ \bibinfo {pages}
  {194301} (\bibinfo {year} {2018})}\BibitemShut {NoStop}%
\bibitem [{\citenamefont {Trainiti}\ \emph {et~al.}(2019)\citenamefont
  {Trainiti}, \citenamefont {Xia}, \citenamefont {Marconi}, \citenamefont
  {Cazzulani}, \citenamefont {Erturk},\ and\ \citenamefont
  {Ruzzene}}]{Ruzzene19}%
  \BibitemOpen
  \bibfield  {author} {\bibinfo {author} {\bibfnamefont {G.}~\bibnamefont
  {Trainiti}}, \bibinfo {author} {\bibfnamefont {Y.}~\bibnamefont {Xia}},
  \bibinfo {author} {\bibfnamefont {J.}~\bibnamefont {Marconi}}, \bibinfo
  {author} {\bibfnamefont {G.}~\bibnamefont {Cazzulani}}, \bibinfo {author}
  {\bibfnamefont {A.}~\bibnamefont {Erturk}}, \ and\ \bibinfo {author}
  {\bibfnamefont {M.}~\bibnamefont {Ruzzene}},\ }\href {\doibase
  10.1103/PhysRevLett.122.124301} {\bibfield  {journal} {\bibinfo  {journal}
  {Phys. Rev. Lett.}\ }\textbf {\bibinfo {volume} {122}},\ \bibinfo {pages}
  {124301} (\bibinfo {year} {2019})}\BibitemShut {NoStop}%
\bibitem [{\citenamefont {Estep}\ \emph {et~al.}(2014)\citenamefont {Estep},
  \citenamefont {Sounas}, \citenamefont {Soric},\ and\ \citenamefont
  {Al{\`u}}}]{Estep14}%
  \BibitemOpen
  \bibfield  {author} {\bibinfo {author} {\bibfnamefont {N.~A.}\ \bibnamefont
  {Estep}}, \bibinfo {author} {\bibfnamefont {D.~L.}\ \bibnamefont {Sounas}},
  \bibinfo {author} {\bibfnamefont {J.}~\bibnamefont {Soric}}, \ and\ \bibinfo
  {author} {\bibfnamefont {A.}~\bibnamefont {Al{\`u}}},\ }\href@noop {}
  {\bibfield  {journal} {\bibinfo  {journal} {Nat. Phys.}\ }\textbf {\bibinfo
  {volume} {10}},\ \bibinfo {pages} {923} (\bibinfo {year} {2014})}\BibitemShut
  {NoStop}%
\bibitem [{\citenamefont {Coulais}\ \emph {et~al.}(2017)\citenamefont
  {Coulais}, \citenamefont {Sounas},\ and\ \citenamefont
  {Al{\`u}}}]{Coulais17}%
  \BibitemOpen
  \bibfield  {author} {\bibinfo {author} {\bibfnamefont {C.}~\bibnamefont
  {Coulais}}, \bibinfo {author} {\bibfnamefont {D.}~\bibnamefont {Sounas}}, \
  and\ \bibinfo {author} {\bibfnamefont {A.}~\bibnamefont {Al{\`u}}},\
  }\href@noop {} {\bibfield  {journal} {\bibinfo  {journal} {Nature (London)}\
  }\textbf {\bibinfo {volume} {542}},\ \bibinfo {pages} {461} (\bibinfo {year}
  {2017})}\BibitemShut {NoStop}%
\bibitem [{\citenamefont {Brandenbourger}\ \emph {et~al.}(2019)\citenamefont
  {Brandenbourger}, \citenamefont {Locsin}, \citenamefont {Lerner},\ and\
  \citenamefont {Coulais}}]{Coulais19}%
  \BibitemOpen
  \bibfield  {author} {\bibinfo {author} {\bibfnamefont {M.}~\bibnamefont
  {Brandenbourger}}, \bibinfo {author} {\bibfnamefont {X.}~\bibnamefont
  {Locsin}}, \bibinfo {author} {\bibfnamefont {E.}~\bibnamefont {Lerner}}, \
  and\ \bibinfo {author} {\bibfnamefont {C.}~\bibnamefont {Coulais}},\
  }\href@noop {} {\bibfield  {journal} {\bibinfo  {journal} {Nat. Commun.}\
  }\textbf {\bibinfo {volume} {10}},\ \bibinfo {pages} {4608} (\bibinfo {year}
  {2019})}\BibitemShut {NoStop}%
\bibitem [{\citenamefont {Liao}\ \emph {et~al.}(2020)\citenamefont {Liao},
  \citenamefont {Irvine},\ and\ \citenamefont {Vaikuntanathan}}]{Liao20}%
  \BibitemOpen
  \bibfield  {author} {\bibinfo {author} {\bibfnamefont {Z.}~\bibnamefont
  {Liao}}, \bibinfo {author} {\bibfnamefont {W.~T.~M.}\ \bibnamefont {Irvine}},
  \ and\ \bibinfo {author} {\bibfnamefont {S.}~\bibnamefont {Vaikuntanathan}},\
  }\href {\doibase 10.1103/PhysRevX.10.021036} {\bibfield  {journal} {\bibinfo
  {journal} {Phys. Rev. X}\ }\textbf {\bibinfo {volume} {10}},\ \bibinfo
  {pages} {021036} (\bibinfo {year} {2020})}\BibitemShut {NoStop}%
\bibitem [{\citenamefont {Flach}\ \emph {et~al.}(2002)\citenamefont {Flach},
  \citenamefont {Zolotaryuk}, \citenamefont {Miroshnichenko},\ and\
  \citenamefont {Fistul}}]{Flach02}%
  \BibitemOpen
  \bibfield  {author} {\bibinfo {author} {\bibfnamefont {S.}~\bibnamefont
  {Flach}}, \bibinfo {author} {\bibfnamefont {Y.}~\bibnamefont {Zolotaryuk}},
  \bibinfo {author} {\bibfnamefont {A.~E.}\ \bibnamefont {Miroshnichenko}}, \
  and\ \bibinfo {author} {\bibfnamefont {M.~V.}\ \bibnamefont {Fistul}},\
  }\href {\doibase 10.1103/PhysRevLett.88.184101} {\bibfield  {journal}
  {\bibinfo  {journal} {Phys. Rev. Lett.}\ }\textbf {\bibinfo {volume} {88}},\
  \bibinfo {pages} {184101} (\bibinfo {year} {2002})}\BibitemShut {NoStop}%
\bibitem [{\citenamefont {Das}\ \emph {et~al.}(2002)\citenamefont {Das},
  \citenamefont {Narayan},\ and\ \citenamefont {Ramaswamy}}]{Souvik02}%
  \BibitemOpen
  \bibfield  {author} {\bibinfo {author} {\bibfnamefont {S.}~\bibnamefont
  {Das}}, \bibinfo {author} {\bibfnamefont {O.}~\bibnamefont {Narayan}}, \ and\
  \bibinfo {author} {\bibfnamefont {S.}~\bibnamefont {Ramaswamy}},\ }\href
  {\doibase 10.1103/PhysRevE.66.050103} {\bibfield  {journal} {\bibinfo
  {journal} {Phys. Rev. E}\ }\textbf {\bibinfo {volume} {66}},\ \bibinfo
  {pages} {050103(R)} (\bibinfo {year} {2002})}\BibitemShut {NoStop}%
\bibitem [{\citenamefont {Ren}\ and\ \citenamefont {Li}(2010)}]{Ren10}%
  \BibitemOpen
  \bibfield  {author} {\bibinfo {author} {\bibfnamefont {J.}~\bibnamefont
  {Ren}}\ and\ \bibinfo {author} {\bibfnamefont {B.}~\bibnamefont {Li}},\
  }\href {\doibase 10.1103/PhysRevE.81.021111} {\bibfield  {journal} {\bibinfo
  {journal} {Phys. Rev. E}\ }\textbf {\bibinfo {volume} {81}},\ \bibinfo
  {pages} {021111} (\bibinfo {year} {2010})}\BibitemShut {NoStop}%
\bibitem [{\citenamefont {Zhu}\ and\ \citenamefont {Fan}(2016)}]{Zhu16}%
  \BibitemOpen
  \bibfield  {author} {\bibinfo {author} {\bibfnamefont {L.}~\bibnamefont
  {Zhu}}\ and\ \bibinfo {author} {\bibfnamefont {S.}~\bibnamefont {Fan}},\
  }\href {\doibase 10.1103/PhysRevLett.117.134303} {\bibfield  {journal}
  {\bibinfo  {journal} {Phys. Rev. Lett.}\ }\textbf {\bibinfo {volume} {117}},\
  \bibinfo {pages} {134303} (\bibinfo {year} {2016})}\BibitemShut {NoStop}%
\bibitem [{\citenamefont {Sabass}(2017)}]{Sabass17}%
  \BibitemOpen
  \bibfield  {author} {\bibinfo {author} {\bibfnamefont {B.}~\bibnamefont
  {Sabass}},\ }\href {\doibase 10.1103/PhysRevE.96.022109} {\bibfield
  {journal} {\bibinfo  {journal} {Phys. Rev. E}\ }\textbf {\bibinfo {volume}
  {96}},\ \bibinfo {pages} {022109} (\bibinfo {year} {2017})}\BibitemShut
  {NoStop}%
\bibitem [{\citenamefont {Onsager}(1931)}]{Onsager31}%
  \BibitemOpen
  \bibfield  {author} {\bibinfo {author} {\bibfnamefont {L.}~\bibnamefont
  {Onsager}},\ }\href {\doibase 10.1103/PhysRev.37.405} {\bibfield  {journal}
  {\bibinfo  {journal} {Phys. Rev.}\ }\textbf {\bibinfo {volume} {37}},\
  \bibinfo {pages} {405} (\bibinfo {year} {1931})}\BibitemShut {NoStop}%
\bibitem [{\citenamefont {Benenti}\ \emph {et~al.}(2017)\citenamefont
  {Benenti}, \citenamefont {Casati}, \citenamefont {Saito},\ and\ \citenamefont
  {Whitney}}]{Benenti17}%
  \BibitemOpen
  \bibfield  {author} {\bibinfo {author} {\bibfnamefont {G.}~\bibnamefont
  {Benenti}}, \bibinfo {author} {\bibfnamefont {G.}~\bibnamefont {Casati}},
  \bibinfo {author} {\bibfnamefont {K.}~\bibnamefont {Saito}}, \ and\ \bibinfo
  {author} {\bibfnamefont {R.~S.}\ \bibnamefont {Whitney}},\ }\href@noop {}
  {\bibfield  {journal} {\bibinfo  {journal} {Phys. Rep.}\ }\textbf {\bibinfo
  {volume} {694}},\ \bibinfo {pages} {1} (\bibinfo {year} {2017})}\BibitemShut
  {NoStop}%
\bibitem [{\citenamefont {Werlang}\ \emph {et~al.}(2014)\citenamefont
  {Werlang}, \citenamefont {Marchiori}, \citenamefont {Cornelio},\ and\
  \citenamefont {Valente}}]{Werlang14}%
  \BibitemOpen
  \bibfield  {author} {\bibinfo {author} {\bibfnamefont {T.}~\bibnamefont
  {Werlang}}, \bibinfo {author} {\bibfnamefont {M.~A.}\ \bibnamefont
  {Marchiori}}, \bibinfo {author} {\bibfnamefont {M.~F.}\ \bibnamefont
  {Cornelio}}, \ and\ \bibinfo {author} {\bibfnamefont {D.}~\bibnamefont
  {Valente}},\ }\href {\doibase 10.1103/PhysRevE.89.062109} {\bibfield
  {journal} {\bibinfo  {journal} {Phys. Rev. E}\ }\textbf {\bibinfo {volume}
  {89}},\ \bibinfo {pages} {062109} (\bibinfo {year} {2014})}\BibitemShut
  {NoStop}%
\bibitem [{\citenamefont {Pereira}(2017{\natexlab{a}})}]{Pereira17b}%
  \BibitemOpen
  \bibfield  {author} {\bibinfo {author} {\bibfnamefont {E.}~\bibnamefont
  {Pereira}},\ }\href {\doibase 10.1103/PhysRevE.96.012114} {\bibfield
  {journal} {\bibinfo  {journal} {Phys. Rev. E}\ }\textbf {\bibinfo {volume}
  {96}},\ \bibinfo {pages} {012114} (\bibinfo {year}
  {2017}{\natexlab{a}})}\BibitemShut {NoStop}%
\bibitem [{\citenamefont {Motz}\ \emph {et~al.}(2018)\citenamefont {Motz},
  \citenamefont {Wiedmann}, \citenamefont {Stockburger},\ and\ \citenamefont
  {Ankerhold}}]{Thomas18}%
  \BibitemOpen
  \bibfield  {author} {\bibinfo {author} {\bibfnamefont {T.}~\bibnamefont
  {Motz}}, \bibinfo {author} {\bibfnamefont {M.}~\bibnamefont {Wiedmann}},
  \bibinfo {author} {\bibfnamefont {J.~T.}\ \bibnamefont {Stockburger}}, \ and\
  \bibinfo {author} {\bibfnamefont {J.}~\bibnamefont {Ankerhold}},\ }\href@noop
  {} {\bibfield  {journal} {\bibinfo  {journal} {New J. Phys.}\ }\textbf
  {\bibinfo {volume} {20}},\ \bibinfo {pages} {113020} (\bibinfo {year}
  {2018})}\BibitemShut {NoStop}%
\bibitem [{\citenamefont {Pereira}(2019{\natexlab{a}})}]{Pereira19}%
  \BibitemOpen
  \bibfield  {author} {\bibinfo {author} {\bibfnamefont {E.}~\bibnamefont
  {Pereira}},\ }\href@noop {} {\bibfield  {journal} {\bibinfo  {journal}
  {Europhys. Lett.}\ }\textbf {\bibinfo {volume} {126}},\ \bibinfo {pages}
  {14001} (\bibinfo {year} {2019}{\natexlab{a}})}\BibitemShut {NoStop}%
\bibitem [{\citenamefont {Pereira}(2019{\natexlab{b}})}]{Pereira19E}%
  \BibitemOpen
  \bibfield  {author} {\bibinfo {author} {\bibfnamefont {E.}~\bibnamefont
  {Pereira}},\ }\href {\doibase 10.1103/PhysRevE.99.032116} {\bibfield
  {journal} {\bibinfo  {journal} {Phys. Rev. E}\ }\textbf {\bibinfo {volume}
  {99}},\ \bibinfo {pages} {032116} (\bibinfo {year}
  {2019}{\natexlab{b}})}\BibitemShut {NoStop}%
\bibitem [{\citenamefont {Riera-Campeny}\ \emph {et~al.}(2019)\citenamefont
  {Riera-Campeny}, \citenamefont {Mehboudi}, \citenamefont {Pons},\ and\
  \citenamefont {Sanpera}}]{Andreu19}%
  \BibitemOpen
  \bibfield  {author} {\bibinfo {author} {\bibfnamefont {A.}~\bibnamefont
  {Riera-Campeny}}, \bibinfo {author} {\bibfnamefont {M.}~\bibnamefont
  {Mehboudi}}, \bibinfo {author} {\bibfnamefont {M.}~\bibnamefont {Pons}}, \
  and\ \bibinfo {author} {\bibfnamefont {A.}~\bibnamefont {Sanpera}},\ }\href
  {\doibase 10.1103/PhysRevE.99.032126} {\bibfield  {journal} {\bibinfo
  {journal} {Phys. Rev. E}\ }\textbf {\bibinfo {volume} {99}},\ \bibinfo
  {pages} {032126} (\bibinfo {year} {2019})}\BibitemShut {NoStop}%
\bibitem [{\citenamefont {Balachandran}\ \emph {et~al.}(2019)\citenamefont
  {Balachandran}, \citenamefont {Benenti}, \citenamefont {Pereira},
  \citenamefont {Casati},\ and\ \citenamefont {Poletti}}]{Balachandran19}%
  \BibitemOpen
  \bibfield  {author} {\bibinfo {author} {\bibfnamefont {V.}~\bibnamefont
  {Balachandran}}, \bibinfo {author} {\bibfnamefont {G.}~\bibnamefont
  {Benenti}}, \bibinfo {author} {\bibfnamefont {E.}~\bibnamefont {Pereira}},
  \bibinfo {author} {\bibfnamefont {G.}~\bibnamefont {Casati}}, \ and\ \bibinfo
  {author} {\bibfnamefont {D.}~\bibnamefont {Poletti}},\ }\href {\doibase
  10.1103/PhysRevE.99.032136} {\bibfield  {journal} {\bibinfo  {journal} {Phys.
  Rev. E}\ }\textbf {\bibinfo {volume} {99}},\ \bibinfo {pages} {032136}
  (\bibinfo {year} {2019})}\BibitemShut {NoStop}%
\bibitem [{\citenamefont {Saito}(2003)}]{Saito03}%
  \BibitemOpen
  \bibfield  {author} {\bibinfo {author} {\bibfnamefont {K.}~\bibnamefont
  {Saito}},\ }\href@noop {} {\bibfield  {journal} {\bibinfo  {journal}
  {Europhys. Lett.}\ }\textbf {\bibinfo {volume} {61}},\ \bibinfo {pages} {34}
  (\bibinfo {year} {2003})}\BibitemShut {NoStop}%
\bibitem [{\citenamefont {Wichterich}\ \emph {et~al.}(2007)\citenamefont
  {Wichterich}, \citenamefont {Henrich}, \citenamefont {Breuer}, \citenamefont
  {Gemmer},\ and\ \citenamefont {Michel}}]{Hannu07}%
  \BibitemOpen
  \bibfield  {author} {\bibinfo {author} {\bibfnamefont {H.}~\bibnamefont
  {Wichterich}}, \bibinfo {author} {\bibfnamefont {M.~J.}\ \bibnamefont
  {Henrich}}, \bibinfo {author} {\bibfnamefont {H.-P.}\ \bibnamefont {Breuer}},
  \bibinfo {author} {\bibfnamefont {J.}~\bibnamefont {Gemmer}}, \ and\ \bibinfo
  {author} {\bibfnamefont {M.}~\bibnamefont {Michel}},\ }\href {\doibase
  10.1103/PhysRevE.76.031115} {\bibfield  {journal} {\bibinfo  {journal} {Phys.
  Rev. E}\ }\textbf {\bibinfo {volume} {76}},\ \bibinfo {pages} {031115}
  (\bibinfo {year} {2007})}\BibitemShut {NoStop}%
\bibitem [{\citenamefont {Prosen}(2011)}]{Prosen11}%
  \BibitemOpen
  \bibfield  {author} {\bibinfo {author} {\bibfnamefont {T.}~\bibnamefont
  {Prosen}},\ }\href {\doibase 10.1103/PhysRevLett.106.217206} {\bibfield
  {journal} {\bibinfo  {journal} {Phys. Rev. Lett.}\ }\textbf {\bibinfo
  {volume} {106}},\ \bibinfo {pages} {217206} (\bibinfo {year}
  {2011})}\BibitemShut {NoStop}%
\bibitem [{\citenamefont {Landi}\ \emph {et~al.}(2014)\citenamefont {Landi},
  \citenamefont {Novais}, \citenamefont {de~Oliveira},\ and\ \citenamefont
  {Karevski}}]{Landi14}%
  \BibitemOpen
  \bibfield  {author} {\bibinfo {author} {\bibfnamefont {G.~T.}\ \bibnamefont
  {Landi}}, \bibinfo {author} {\bibfnamefont {E.}~\bibnamefont {Novais}},
  \bibinfo {author} {\bibfnamefont {M.~J.}\ \bibnamefont {de~Oliveira}}, \ and\
  \bibinfo {author} {\bibfnamefont {D.}~\bibnamefont {Karevski}},\ }\href
  {\doibase 10.1103/PhysRevE.90.042142} {\bibfield  {journal} {\bibinfo
  {journal} {Phys. Rev. E}\ }\textbf {\bibinfo {volume} {90}},\ \bibinfo
  {pages} {042142} (\bibinfo {year} {2014})}\BibitemShut {NoStop}%
\bibitem [{\citenamefont {Lenar\ifmmode \check{c}\else
  \v{c}\fi{}i\ifmmode~\check{c}\else \v{c}\fi{}}\ and\ \citenamefont
  {Prosen}(2015)}]{Zala15}%
  \BibitemOpen
  \bibfield  {author} {\bibinfo {author} {\bibfnamefont {Z.}~\bibnamefont
  {Lenar\ifmmode \check{c}\else \v{c}\fi{}i\ifmmode~\check{c}\else
  \v{c}\fi{}}}\ and\ \bibinfo {author} {\bibfnamefont {T.}~\bibnamefont
  {Prosen}},\ }\href {\doibase 10.1103/PhysRevE.91.030103} {\bibfield
  {journal} {\bibinfo  {journal} {Phys. Rev. E}\ }\textbf {\bibinfo {volume}
  {91}},\ \bibinfo {pages} {030103(R)} (\bibinfo {year} {2015})}\BibitemShut
  {NoStop}%
\bibitem [{\citenamefont {Pereira}(2017{\natexlab{b}})}]{Pereira17a}%
  \BibitemOpen
  \bibfield  {author} {\bibinfo {author} {\bibfnamefont {E.}~\bibnamefont
  {Pereira}},\ }\href {\doibase 10.1103/PhysRevE.95.030104} {\bibfield
  {journal} {\bibinfo  {journal} {Phys. Rev. E}\ }\textbf {\bibinfo {volume}
  {95}},\ \bibinfo {pages} {030104(R)} (\bibinfo {year}
  {2017}{\natexlab{b}})}\BibitemShut {NoStop}%
\bibitem [{\citenamefont {Pereira}(2018)}]{Pereira18}%
  \BibitemOpen
  \bibfield  {author} {\bibinfo {author} {\bibfnamefont {E.}~\bibnamefont
  {Pereira}},\ }\href {\doibase 10.1103/PhysRevE.97.022115} {\bibfield
  {journal} {\bibinfo  {journal} {Phys. Rev. E}\ }\textbf {\bibinfo {volume}
  {97}},\ \bibinfo {pages} {022115} (\bibinfo {year} {2018})}\BibitemShut
  {NoStop}%
\bibitem [{\citenamefont {Schuab}\ \emph {et~al.}(2016)\citenamefont {Schuab},
  \citenamefont {Pereira},\ and\ \citenamefont {Landi}}]{Lucas16}%
  \BibitemOpen
  \bibfield  {author} {\bibinfo {author} {\bibfnamefont {L.}~\bibnamefont
  {Schuab}}, \bibinfo {author} {\bibfnamefont {E.}~\bibnamefont {Pereira}}, \
  and\ \bibinfo {author} {\bibfnamefont {G.~T.}\ \bibnamefont {Landi}},\ }\href
  {\doibase 10.1103/PhysRevE.94.042122} {\bibfield  {journal} {\bibinfo
  {journal} {Phys. Rev. E}\ }\textbf {\bibinfo {volume} {94}},\ \bibinfo
  {pages} {042122} (\bibinfo {year} {2016})}\BibitemShut {NoStop}%
\bibitem [{\citenamefont {Balachandran}\ \emph {et~al.}(2018)\citenamefont
  {Balachandran}, \citenamefont {Benenti}, \citenamefont {Pereira},
  \citenamefont {Casati},\ and\ \citenamefont {Poletti}}]{Balachandran18}%
  \BibitemOpen
  \bibfield  {author} {\bibinfo {author} {\bibfnamefont {V.}~\bibnamefont
  {Balachandran}}, \bibinfo {author} {\bibfnamefont {G.}~\bibnamefont
  {Benenti}}, \bibinfo {author} {\bibfnamefont {E.}~\bibnamefont {Pereira}},
  \bibinfo {author} {\bibfnamefont {G.}~\bibnamefont {Casati}}, \ and\ \bibinfo
  {author} {\bibfnamefont {D.}~\bibnamefont {Poletti}},\ }\href {\doibase
  10.1103/PhysRevLett.120.200603} {\bibfield  {journal} {\bibinfo  {journal}
  {Phys. Rev. Lett.}\ }\textbf {\bibinfo {volume} {120}},\ \bibinfo {pages}
  {200603} (\bibinfo {year} {2018})}\BibitemShut {NoStop}%
\bibitem [{\citenamefont {Malz}\ and\ \citenamefont
  {Nunnenkamp}(2018)}]{Daniel18}%
  \BibitemOpen
  \bibfield  {author} {\bibinfo {author} {\bibfnamefont {D.}~\bibnamefont
  {Malz}}\ and\ \bibinfo {author} {\bibfnamefont {A.}~\bibnamefont
  {Nunnenkamp}},\ }\href {\doibase 10.1103/PhysRevB.97.165308} {\bibfield
  {journal} {\bibinfo  {journal} {Phys. Rev. B}\ }\textbf {\bibinfo {volume}
  {97}},\ \bibinfo {pages} {165308} (\bibinfo {year} {2018})}\BibitemShut
  {NoStop}%
\bibitem [{\citenamefont {Hovhannisyan}\ and\ \citenamefont
  {Imparato}(2019)}]{Karen19}%
  \BibitemOpen
  \bibfield  {author} {\bibinfo {author} {\bibfnamefont {K.~V.}\ \bibnamefont
  {Hovhannisyan}}\ and\ \bibinfo {author} {\bibfnamefont {A.}~\bibnamefont
  {Imparato}},\ }\href@noop {} {\bibfield  {journal} {\bibinfo  {journal} {New
  J. Phys.}\ }\textbf {\bibinfo {volume} {21}},\ \bibinfo {pages} {052001}
  (\bibinfo {year} {2019})}\BibitemShut {NoStop}%
\bibitem [{\citenamefont {Mascarenhas}\ \emph {et~al.}(2019)\citenamefont
  {Mascarenhas}, \citenamefont {Damanet}, \citenamefont {Flannigan},
  \citenamefont {Tagliacozzo}, \citenamefont {Daley}, \citenamefont {Goold},\
  and\ \citenamefont {de~Vega}}]{Eduardo19}%
  \BibitemOpen
  \bibfield  {author} {\bibinfo {author} {\bibfnamefont {E.}~\bibnamefont
  {Mascarenhas}}, \bibinfo {author} {\bibfnamefont {F.}~\bibnamefont
  {Damanet}}, \bibinfo {author} {\bibfnamefont {S.}~\bibnamefont {Flannigan}},
  \bibinfo {author} {\bibfnamefont {L.}~\bibnamefont {Tagliacozzo}}, \bibinfo
  {author} {\bibfnamefont {A.~J.}\ \bibnamefont {Daley}}, \bibinfo {author}
  {\bibfnamefont {J.}~\bibnamefont {Goold}}, \ and\ \bibinfo {author}
  {\bibfnamefont {I.}~\bibnamefont {de~Vega}},\ }\href {\doibase
  10.1103/PhysRevB.99.245134} {\bibfield  {journal} {\bibinfo  {journal} {Phys.
  Rev. B}\ }\textbf {\bibinfo {volume} {99}},\ \bibinfo {pages} {245134}
  (\bibinfo {year} {2019})}\BibitemShut {NoStop}%
\bibitem [{\citenamefont {Damanet}\ \emph {et~al.}(2019)\citenamefont
  {Damanet}, \citenamefont {Mascarenhas}, \citenamefont {Pekker},\ and\
  \citenamefont {Daley}}]{Damanet19}%
  \BibitemOpen
  \bibfield  {author} {\bibinfo {author} {\bibfnamefont {F.}~\bibnamefont
  {Damanet}}, \bibinfo {author} {\bibfnamefont {E.}~\bibnamefont
  {Mascarenhas}}, \bibinfo {author} {\bibfnamefont {D.}~\bibnamefont {Pekker}},
  \ and\ \bibinfo {author} {\bibfnamefont {A.~J.}\ \bibnamefont {Daley}},\
  }\href {\doibase 10.1103/PhysRevLett.123.180402} {\bibfield  {journal}
  {\bibinfo  {journal} {Phys. Rev. Lett.}\ }\textbf {\bibinfo {volume} {123}},\
  \bibinfo {pages} {180402} (\bibinfo {year} {2019})}\BibitemShut {NoStop}%
\bibitem [{\citenamefont {Metelmann}\ and\ \citenamefont
  {Clerk}(2015)}]{Metelmann15}%
  \BibitemOpen
  \bibfield  {author} {\bibinfo {author} {\bibfnamefont {A.}~\bibnamefont
  {Metelmann}}\ and\ \bibinfo {author} {\bibfnamefont {A.~A.}\ \bibnamefont
  {Clerk}},\ }\href {\doibase 10.1103/PhysRevX.5.021025} {\bibfield  {journal}
  {\bibinfo  {journal} {Phys. Rev. X}\ }\textbf {\bibinfo {volume} {5}},\
  \bibinfo {pages} {021025} (\bibinfo {year} {2015})}\BibitemShut {NoStop}%
\bibitem [{\citenamefont {Lodahl}\ \emph {et~al.}(2017)\citenamefont {Lodahl},
  \citenamefont {Mahmoodian}, \citenamefont {Stobbe}, \citenamefont
  {Rauschenbeutel}, \citenamefont {Schneeweiss}, \citenamefont {Volz},
  \citenamefont {Pichler},\ and\ \citenamefont {Zoller}}]{Lodahl17}%
  \BibitemOpen
  \bibfield  {author} {\bibinfo {author} {\bibfnamefont {P.}~\bibnamefont
  {Lodahl}}, \bibinfo {author} {\bibfnamefont {S.}~\bibnamefont {Mahmoodian}},
  \bibinfo {author} {\bibfnamefont {S.}~\bibnamefont {Stobbe}}, \bibinfo
  {author} {\bibfnamefont {A.}~\bibnamefont {Rauschenbeutel}}, \bibinfo
  {author} {\bibfnamefont {P.}~\bibnamefont {Schneeweiss}}, \bibinfo {author}
  {\bibfnamefont {J.}~\bibnamefont {Volz}}, \bibinfo {author} {\bibfnamefont
  {H.}~\bibnamefont {Pichler}}, \ and\ \bibinfo {author} {\bibfnamefont
  {P.}~\bibnamefont {Zoller}},\ }\href@noop {} {\bibfield  {journal} {\bibinfo
  {journal} {Nature (London)}\ }\textbf {\bibinfo {volume} {541}},\ \bibinfo
  {pages} {473} (\bibinfo {year} {2017})}\BibitemShut {NoStop}%
\bibitem [{\citenamefont {Keck}\ \emph {et~al.}(2018)\citenamefont {Keck},
  \citenamefont {Rossini},\ and\ \citenamefont {Fazio}}]{Keck18}%
  \BibitemOpen
  \bibfield  {author} {\bibinfo {author} {\bibfnamefont {M.}~\bibnamefont
  {Keck}}, \bibinfo {author} {\bibfnamefont {D.}~\bibnamefont {Rossini}}, \
  and\ \bibinfo {author} {\bibfnamefont {R.}~\bibnamefont {Fazio}},\ }\href
  {\doibase 10.1103/PhysRevA.98.053812} {\bibfield  {journal} {\bibinfo
  {journal} {Phys. Rev. A}\ }\textbf {\bibinfo {volume} {98}},\ \bibinfo
  {pages} {053812} (\bibinfo {year} {2018})}\BibitemShut {NoStop}%
\bibitem [{\citenamefont {Lange}\ \emph {et~al.}(2018)\citenamefont {Lange},
  \citenamefont {Lenar\ifmmode \check{c}\else
  \v{c}\fi{}i\ifmmode~\check{c}\else \v{c}\fi{}},\ and\ \citenamefont
  {Rosch}}]{Zala18a}%
  \BibitemOpen
  \bibfield  {author} {\bibinfo {author} {\bibfnamefont {F.}~\bibnamefont
  {Lange}}, \bibinfo {author} {\bibfnamefont {Z.}~\bibnamefont {Lenar\ifmmode
  \check{c}\else \v{c}\fi{}i\ifmmode~\check{c}\else \v{c}\fi{}}}, \ and\
  \bibinfo {author} {\bibfnamefont {A.}~\bibnamefont {Rosch}},\ }\href
  {\doibase 10.1103/PhysRevB.97.165138} {\bibfield  {journal} {\bibinfo
  {journal} {Phys. Rev. B}\ }\textbf {\bibinfo {volume} {97}},\ \bibinfo
  {pages} {165138} (\bibinfo {year} {2018})}\BibitemShut {NoStop}%
\bibitem [{\citenamefont {Birkholz}\ and\ \citenamefont
  {Meden}(2008)}]{Birkholz08}%
  \BibitemOpen
  \bibfield  {author} {\bibinfo {author} {\bibfnamefont {J.}~\bibnamefont
  {Birkholz}}\ and\ \bibinfo {author} {\bibfnamefont {V.}~\bibnamefont
  {Meden}},\ }\href@noop {} {\bibfield  {journal} {\bibinfo  {journal} {J.
  Condens. Matter Phys.}\ }\textbf {\bibinfo {volume} {20}},\ \bibinfo {pages}
  {085226} (\bibinfo {year} {2008})}\BibitemShut {NoStop}%
\bibitem [{\citenamefont {Nakajima}\ \emph {et~al.}(2018)\citenamefont
  {Nakajima}, \citenamefont {Delbecq}, \citenamefont {Otsuka}, \citenamefont
  {Amaha}, \citenamefont {Yoneda}, \citenamefont {Noiri}, \citenamefont
  {Takeda}, \citenamefont {Allison}, \citenamefont {Ludwig}, \citenamefont
  {Wieck}, \citenamefont {Hu}, \citenamefont {Nori},\ and\ \citenamefont
  {Tarucha}}]{Nakajima18}%
  \BibitemOpen
  \bibfield  {author} {\bibinfo {author} {\bibfnamefont {T.}~\bibnamefont
  {Nakajima}}, \bibinfo {author} {\bibfnamefont {M.~R.}\ \bibnamefont
  {Delbecq}}, \bibinfo {author} {\bibfnamefont {T.}~\bibnamefont {Otsuka}},
  \bibinfo {author} {\bibfnamefont {S.}~\bibnamefont {Amaha}}, \bibinfo
  {author} {\bibfnamefont {J.}~\bibnamefont {Yoneda}}, \bibinfo {author}
  {\bibfnamefont {A.}~\bibnamefont {Noiri}}, \bibinfo {author} {\bibfnamefont
  {K.}~\bibnamefont {Takeda}}, \bibinfo {author} {\bibfnamefont
  {G.}~\bibnamefont {Allison}}, \bibinfo {author} {\bibfnamefont
  {A.}~\bibnamefont {Ludwig}}, \bibinfo {author} {\bibfnamefont {A.~D.}\
  \bibnamefont {Wieck}}, \bibinfo {author} {\bibfnamefont {X.}~\bibnamefont
  {Hu}}, \bibinfo {author} {\bibfnamefont {F.}~\bibnamefont {Nori}}, \ and\
  \bibinfo {author} {\bibfnamefont {S.}~\bibnamefont {Tarucha}},\ }\href@noop
  {} {\bibfield  {journal} {\bibinfo  {journal} {Nat. Commun.}\ }\textbf
  {\bibinfo {volume} {9}},\ \bibinfo {pages} {2133} (\bibinfo {year}
  {2018})}\BibitemShut {NoStop}%
\bibitem [{\citenamefont {Vidmar}\ and\ \citenamefont {Rigol}()}]{Vidmar16}%
  \BibitemOpen
  \bibfield  {author} {\bibinfo {author} {\bibfnamefont {L.}~\bibnamefont
  {Vidmar}}\ and\ \bibinfo {author} {\bibfnamefont {M.}~\bibnamefont {Rigol}},\
  }\href@noop {} {\bibinfo  {journal} {J. Stat. Mech. (2016) 064007}\
  }\BibitemShut {NoStop}%
\bibitem [{\citenamefont {Lange}\ \emph {et~al.}(2017)\citenamefont {Lange},
  \citenamefont {Lenar{\v{c}}i{\v{c}}},\ and\ \citenamefont {Rosch}}]{Zala17}%
  \BibitemOpen
\bibfield  {journal} {  }\bibfield  {author} {\bibinfo {author} {\bibfnamefont
  {F.}~\bibnamefont {Lange}}, \bibinfo {author} {\bibfnamefont
  {Z.}~\bibnamefont {Lenar{\v{c}}i{\v{c}}}}, \ and\ \bibinfo {author}
  {\bibfnamefont {A.}~\bibnamefont {Rosch}},\ }\href@noop {} {\bibfield
  {journal} {\bibinfo  {journal} {Nat. Commun.}\ }\textbf {\bibinfo {volume}
  {8}},\ \bibinfo {pages} {15767} (\bibinfo {year} {2017})}\BibitemShut
  {NoStop}%
\bibitem [{\citenamefont {Lenar\ifmmode \check{c}\else
  \v{c}\fi{}i\ifmmode~\check{c}\else \v{c}\fi{}}\ \emph
  {et~al.}(2018)\citenamefont {Lenar\ifmmode \check{c}\else
  \v{c}\fi{}i\ifmmode~\check{c}\else \v{c}\fi{}}, \citenamefont {Lange},\ and\
  \citenamefont {Rosch}}]{Zala18b}%
  \BibitemOpen
  \bibfield  {author} {\bibinfo {author} {\bibfnamefont {Z.}~\bibnamefont
  {Lenar\ifmmode \check{c}\else \v{c}\fi{}i\ifmmode~\check{c}\else
  \v{c}\fi{}}}, \bibinfo {author} {\bibfnamefont {F.}~\bibnamefont {Lange}}, \
  and\ \bibinfo {author} {\bibfnamefont {A.}~\bibnamefont {Rosch}},\ }\href
  {\doibase 10.1103/PhysRevB.97.024302} {\bibfield  {journal} {\bibinfo
  {journal} {Phys. Rev. B}\ }\textbf {\bibinfo {volume} {97}},\ \bibinfo
  {pages} {024302} (\bibinfo {year} {2018})}\BibitemShut {NoStop}%
\bibitem [{\citenamefont {Liu}(2014)}]{Liu14}%
  \BibitemOpen
  \bibfield  {author} {\bibinfo {author} {\bibfnamefont {F.}~\bibnamefont
  {Liu}},\ }\href {\doibase 10.1103/PhysRevE.89.042122} {\bibfield  {journal}
  {\bibinfo  {journal} {Phys. Rev. E}\ }\textbf {\bibinfo {volume} {89}},\
  \bibinfo {pages} {042122} (\bibinfo {year} {2014})}\BibitemShut {NoStop}%
\bibitem [{\citenamefont {Gong}\ \emph {et~al.}(2016)\citenamefont {Gong},
  \citenamefont {Ashida},\ and\ \citenamefont {Ueda}}]{Gong16}%
  \BibitemOpen
  \bibfield  {author} {\bibinfo {author} {\bibfnamefont {Z.}~\bibnamefont
  {Gong}}, \bibinfo {author} {\bibfnamefont {Y.}~\bibnamefont {Ashida}}, \ and\
  \bibinfo {author} {\bibfnamefont {M.}~\bibnamefont {Ueda}},\ }\href {\doibase
  10.1103/PhysRevA.94.012107} {\bibfield  {journal} {\bibinfo  {journal} {Phys.
  Rev. A}\ }\textbf {\bibinfo {volume} {94}},\ \bibinfo {pages} {012107}
  (\bibinfo {year} {2016})}\BibitemShut {NoStop}%
\bibitem [{\citenamefont {Gebauer}\ and\ \citenamefont
  {Car}(2004)}]{Gebauer04}%
  \BibitemOpen
  \bibfield  {author} {\bibinfo {author} {\bibfnamefont {R.}~\bibnamefont
  {Gebauer}}\ and\ \bibinfo {author} {\bibfnamefont {R.}~\bibnamefont {Car}},\
  }\href {\doibase 10.1103/PhysRevLett.93.160404} {\bibfield  {journal}
  {\bibinfo  {journal} {Phys. Rev. Lett.}\ }\textbf {\bibinfo {volume} {93}},\
  \bibinfo {pages} {160404} (\bibinfo {year} {2004})}\BibitemShut {NoStop}%
\bibitem [{\citenamefont {Bodor}\ and\ \citenamefont
  {Di\'osi}(2006)}]{Bodor06}%
  \BibitemOpen
  \bibfield  {author} {\bibinfo {author} {\bibfnamefont {A.}~\bibnamefont
  {Bodor}}\ and\ \bibinfo {author} {\bibfnamefont {L.}~\bibnamefont
  {Di\'osi}},\ }\href {\doibase 10.1103/PhysRevA.73.064101} {\bibfield
  {journal} {\bibinfo  {journal} {Phys. Rev. A}\ }\textbf {\bibinfo {volume}
  {73}},\ \bibinfo {pages} {064101} (\bibinfo {year} {2006})}\BibitemShut
  {NoStop}%
\bibitem [{\citenamefont {de~Andrada~e Silva}(1992)}]{Erasmo92}%
  \BibitemOpen
  \bibfield  {author} {\bibinfo {author} {\bibfnamefont {E.~A.}\ \bibnamefont
  {de~Andrada~e Silva}},\ }\href@noop {} {\bibfield  {journal} {\bibinfo
  {journal} {Am. J. Phys.}\ }\textbf {\bibinfo {volume} {60}},\ \bibinfo
  {pages} {753} (\bibinfo {year} {1992})}\BibitemShut {NoStop}%
\bibitem [{\citenamefont {Mahan}(2000)}]{Mahan00}%
  \BibitemOpen
  \bibfield  {author} {\bibinfo {author} {\bibfnamefont {G.~D.}\ \bibnamefont
  {Mahan}},\ }\href@noop {} {\emph {\bibinfo {title} {Many-particle physics}}}\
  (\bibinfo  {publisher} {Kluwer Academic / Plenum Publishers, New York},\
  \bibinfo {year} {2000})\BibitemShut {NoStop}%
\bibitem [{\citenamefont {Rikken}\ \emph {et~al.}(2001)\citenamefont {Rikken},
  \citenamefont {F\"olling},\ and\ \citenamefont {Wyder}}]{Rikken01}%
  \BibitemOpen
  \bibfield  {author} {\bibinfo {author} {\bibfnamefont {G.~L. J.~A.}\
  \bibnamefont {Rikken}}, \bibinfo {author} {\bibfnamefont {J.}~\bibnamefont
  {F\"olling}}, \ and\ \bibinfo {author} {\bibfnamefont {P.}~\bibnamefont
  {Wyder}},\ }\href {\doibase 10.1103/PhysRevLett.87.236602} {\bibfield
  {journal} {\bibinfo  {journal} {Phys. Rev. Lett.}\ }\textbf {\bibinfo
  {volume} {87}},\ \bibinfo {pages} {236602} (\bibinfo {year}
  {2001})}\BibitemShut {NoStop}%
\bibitem [{\citenamefont {Morimoto}\ and\ \citenamefont
  {Nagaosa}(2016{\natexlab{b}})}]{Morimoto16L}%
  \BibitemOpen
  \bibfield  {author} {\bibinfo {author} {\bibfnamefont {T.}~\bibnamefont
  {Morimoto}}\ and\ \bibinfo {author} {\bibfnamefont {N.}~\bibnamefont
  {Nagaosa}},\ }\href {\doibase 10.1103/PhysRevLett.117.146603} {\bibfield
  {journal} {\bibinfo  {journal} {Phys. Rev. Lett.}\ }\textbf {\bibinfo
  {volume} {117}},\ \bibinfo {pages} {146603} (\bibinfo {year}
  {2016}{\natexlab{b}})}\BibitemShut {NoStop}%
\bibitem [{\citenamefont {Ideue}\ \emph {et~al.}(2017)\citenamefont {Ideue},
  \citenamefont {Hamamoto}, \citenamefont {Koshikawa}, \citenamefont {Ezawa},
  \citenamefont {Shimizu}, \citenamefont {Kaneko}, \citenamefont {Tokura},
  \citenamefont {Nagaosa},\ and\ \citenamefont {Iwasa}}]{Ideue17}%
  \BibitemOpen
  \bibfield  {author} {\bibinfo {author} {\bibfnamefont {T.}~\bibnamefont
  {Ideue}}, \bibinfo {author} {\bibfnamefont {K.}~\bibnamefont {Hamamoto}},
  \bibinfo {author} {\bibfnamefont {S.}~\bibnamefont {Koshikawa}}, \bibinfo
  {author} {\bibfnamefont {M.}~\bibnamefont {Ezawa}}, \bibinfo {author}
  {\bibfnamefont {S.}~\bibnamefont {Shimizu}}, \bibinfo {author} {\bibfnamefont
  {Y.}~\bibnamefont {Kaneko}}, \bibinfo {author} {\bibfnamefont
  {Y.}~\bibnamefont {Tokura}}, \bibinfo {author} {\bibfnamefont
  {N.}~\bibnamefont {Nagaosa}}, \ and\ \bibinfo {author} {\bibfnamefont
  {Y.}~\bibnamefont {Iwasa}},\ }\href@noop {} {\bibfield  {journal} {\bibinfo
  {journal} {Nat. Phys.}\ }\textbf {\bibinfo {volume} {13}},\ \bibinfo {pages}
  {578} (\bibinfo {year} {2017})}\BibitemShut {NoStop}%
\bibitem [{\citenamefont {St\ifmmode~\check{r}\else \v{r}\fi{}eda}\ and\
  \citenamefont {\ifmmode~\check{S}\else \v{S}\fi{}eba}(2003)}]{Streda03}%
  \BibitemOpen
  \bibfield  {author} {\bibinfo {author} {\bibfnamefont {P.}~\bibnamefont
  {St\ifmmode~\check{r}\else \v{r}\fi{}eda}}\ and\ \bibinfo {author}
  {\bibfnamefont {P.}~\bibnamefont {\ifmmode~\check{S}\else \v{S}\fi{}eba}},\
  }\href {\doibase 10.1103/PhysRevLett.90.256601} {\bibfield  {journal}
  {\bibinfo  {journal} {Phys. Rev. Lett.}\ }\textbf {\bibinfo {volume} {90}},\
  \bibinfo {pages} {256601} (\bibinfo {year} {2003})}\BibitemShut {NoStop}%
\bibitem [{\citenamefont {Nitta}\ \emph {et~al.}(1997)\citenamefont {Nitta},
  \citenamefont {Akazaki}, \citenamefont {Takayanagi},\ and\ \citenamefont
  {Enoki}}]{Nitta97}%
  \BibitemOpen
  \bibfield  {author} {\bibinfo {author} {\bibfnamefont {J.}~\bibnamefont
  {Nitta}}, \bibinfo {author} {\bibfnamefont {T.}~\bibnamefont {Akazaki}},
  \bibinfo {author} {\bibfnamefont {H.}~\bibnamefont {Takayanagi}}, \ and\
  \bibinfo {author} {\bibfnamefont {T.}~\bibnamefont {Enoki}},\ }\href
  {\doibase 10.1103/PhysRevLett.78.1335} {\bibfield  {journal} {\bibinfo
  {journal} {Phys. Rev. Lett.}\ }\textbf {\bibinfo {volume} {78}},\ \bibinfo
  {pages} {1335} (\bibinfo {year} {1997})}\BibitemShut {NoStop}%
\bibitem [{\citenamefont {Grundler}(2000)}]{Dirk00}%
  \BibitemOpen
  \bibfield  {author} {\bibinfo {author} {\bibfnamefont {D.}~\bibnamefont
  {Grundler}},\ }\href {\doibase 10.1103/PhysRevLett.84.6074} {\bibfield
  {journal} {\bibinfo  {journal} {Phys. Rev. Lett.}\ }\textbf {\bibinfo
  {volume} {84}},\ \bibinfo {pages} {6074} (\bibinfo {year}
  {2000})}\BibitemShut {NoStop}%
\bibitem [{\citenamefont {Scarlino}\ \emph {et~al.}(2014)\citenamefont
  {Scarlino}, \citenamefont {Kawakami}, \citenamefont {Stano}, \citenamefont
  {Shafiei}, \citenamefont {Reichl}, \citenamefont {Wegscheider},\ and\
  \citenamefont {Vandersypen}}]{Vandersypen14}%
  \BibitemOpen
  \bibfield  {author} {\bibinfo {author} {\bibfnamefont {P.}~\bibnamefont
  {Scarlino}}, \bibinfo {author} {\bibfnamefont {E.}~\bibnamefont {Kawakami}},
  \bibinfo {author} {\bibfnamefont {P.}~\bibnamefont {Stano}}, \bibinfo
  {author} {\bibfnamefont {M.}~\bibnamefont {Shafiei}}, \bibinfo {author}
  {\bibfnamefont {C.}~\bibnamefont {Reichl}}, \bibinfo {author} {\bibfnamefont
  {W.}~\bibnamefont {Wegscheider}}, \ and\ \bibinfo {author} {\bibfnamefont
  {L.~M.~K.}\ \bibnamefont {Vandersypen}},\ }\href {\doibase
  10.1103/PhysRevLett.113.256802} {\bibfield  {journal} {\bibinfo  {journal}
  {Phys. Rev. Lett.}\ }\textbf {\bibinfo {volume} {113}},\ \bibinfo {pages}
  {256802} (\bibinfo {year} {2014})}\BibitemShut {NoStop}%
\end{thebibliography}%

\end{document}